%% file: 3BF.tex
\documentclass[floatfix,
 aps, prc,     % aps % [prl],[pra],[prb],[prc],[prd],[pre],[prstab],[prstper],[rmp]
 showpacs, % display PACS
 showkeys, % display keywords
 superscriptaddress, % affiliations with superscripts
 reprint, % reprint: appearance close to chosen journal % preprint: for one column
 nofootinbib % at end of page %footinbib: footnote in bibliography, 
 ,onecolumn% onecolumn, twocolumn,
]{revtex4-1}

\usepackage[english]{babel}
\usepackage[T1]{fontenc}
\usepackage[utf8]{inputenc} %latin9
\usepackage{amsmath} % math mode
\usepackage{amssymb, bbm} % mathematica symbols; unit matrix \mathbbm{1}
\usepackage{amstext} % text in math mode with \text{}
\usepackage{mathrsfs} % curved letter, e.g. mathscr{L}
\usepackage{booktabs} \cmidrulekern=.3em % default .5em % nice tables; e.g. \toprule, \midrule, \bottomrule, \cmidrule(lr){1-3}, \addlinespace[3pt]
\usepackage{array} % for tables; e.g. math mode in a column: \begin{tabular}{|c|>{\(}c<{\)}|c|c|c|}
\usepackage{graphicx} \graphicspath{ {./files/}{./figures/} } % \graphicspath{ {./figures/}{./graphs/} }
\usepackage{xcolor} \definecolor{lblue}{RGB}{70,126,185}
\usepackage[pdftex,pdfpagelabels]{hyperref} % hyperfootnotes=false
\hypersetup{
   colorlinks=true, linktocpage=true, urlcolor=lblue, linkcolor=lblue, citecolor=lblue,
   pdftitle={Density-dependent effective baryon-baryon interaction from chiral three-baryon forces},
   pdfauthor={Stefan Petschauer}
}
\usepackage{afterpage}
\usepackage{overpic}

\newcommand{\Ps}{P^{(\sigma)}}

\newcommand{\m}{\ensuremath{m_\pi}}
\DeclareMathOperator{\tr}{tr}

\newcommand\numberthis{\addtocounter{equation}{1}\tag{\theequation}}
\newcommand\undefcolumntype[1]{\expandafter\let\csname NC@find@#1\endcsname\relax}

\makeatletter
\protected\def\lc{C}
\protected\def\ld{D}
\protected\def\lC{H}
\newcommand*{\balancecolsandclearpage}{%
  \close@column@grid
  \cleardoublepage
  \twocolumngrid
}
\protected\def\fig{\@ifstar\@fig\@@fig} \def\@fig#1{\ref{#1}} \def\@@fig#1{Fig.~\ref{#1}}
\protected\def\tab{\@ifstar\@tab\@@tab} \def\@tab#1{\ref{#1}} \def\@@tab#1{Tab.~\ref{#1}}
\protected\def\eq{\@ifstar\@eq\@@eq} \def\@eq#1{\eqref{#1}} \def\@@eq#1{Eq.~\eqref{#1}}
\protected\def\ch{\@ifstar\@ch\@@ch} \def\@ch#1{\ref{#1}} \def\@@ch#1{Ch.~\ref{#1}}
\protected\def\sect{\@ifstar\@sect\@@sect} \def\@sect#1{\ref{#1}} \def\@@sect#1{Sec.~\ref{#1}}
\protected\def\ssect{\@ifstar\@ssect\@@ssect} \def\@ssect#1{\ref{#1}} \def\@@ssect#1{Subsec.~\ref{#1}}
\protected\def\app{\@ifstar\@app\@@app} \def\@app#1{\cite{#1}} \def\@@app#1{Appendix~\ref{#1}}
\protected\def\ct{\@ifstar\@ct\@@ct} \def\@ct#1{\cite{#1}} \def\@@ct#1{Ref.~\cite{#1}}
\protected\def\figs{\@ifstar\@figs\@@figs} \def\@figs#1{\ref{#1}} \def\@@figs#1{Figs.~\ref{#1}}
\protected\def\tabs{\@ifstar\@tabs\@@tabs} \def\@tabs#1{\ref{#1}} \def\@@tabs#1{Tabs.~\ref{#1}}
\protected\def\eqs{\@ifstar\@eqs\@@eqs} \def\@eqs#1{\eqref{#1}} \def\@@eqs#1{Eqs.~\eqref{#1}}
\protected\def\chs{\@ifstar\@chs\@@chs} \def\@chs#1{\ref{#1}} \def\@@chs#1{Chs.~\ref{#1}}
\protected\def\sects{\@ifstar\@sects\@@sects} \def\@sects#1{\ref{#1}} \def\@@sects#1{Secs.~\ref{#1}}
\protected\def\ssects{\@ifstar\@ssects\@@ssects} \def\@ssects#1{\ref{#1}} \def\@@ssects#1{Subsecs.~\ref{#1}}
\protected\def\apps{\@ifstar\@apps\@@apps} \def\@apps#1{\cite{#1}} \def\@@apps#1{Appendices~\ref{#1}}
\protected\def\cts{\@ifstar\@cts\@@cts} \def\@cts#1{\cite{#1}} \def\@@cts#1{Refs.~\cite{#1}}
\makeatother

\newcommand{\ie}{i.e., }

\newcommand{\eg}{e.g., }

\newcommand{\cf}{cf.\ }

\newcommand{\cheft}{\(\chi\)EFT}

\begin{document}

\title{Density-dependent effective baryon-baryon interaction from chiral three-baryon forces}

\author{Stefan~Petschauer}
   \email{stefan.petschauer@ph.tum.de}
   \affiliation{Physik Department, Technische Universit\"at M\"unchen, D-85747 Garching, Germany}
   %% \thanks{note}
\author{Johann~Haidenbauer}
   \affiliation{Institute for Advanced Simulation and J\"ulich Center for Hadron Physics,\\
                Institut f\"ur Kernphysik, Forschungszentrum J\"ulich, D-52425 J\"ulich, Germany}
\author{Norbert~Kaiser}
   \affiliation{Physik Department, Technische Universit\"at M\"unchen, D-85747 Garching, Germany}
\author{Ulf-G.~Mei\ss{}ner}
   \affiliation{Institute for Advanced Simulation and J\"ulich Center for Hadron Physics,\\
                Institut f\"ur Kernphysik, Forschungszentrum J\"ulich, D-52425 J\"ulich, Germany}
   \affiliation{Helmholtz-Institut f\"ur Strahlen- und Kernphysik and Bethe Center\\
                for Theoretical Physics, Universit\"at Bonn, D-53115 Bonn, Germany}
\author{Wolfram~Weise}
   \affiliation{Physik Department, Technische Universit\"at M\"unchen, D-85747 Garching, Germany}

\date{\today}

\begin{abstract}
A density-dependent effective potential for the baryon-baryon interaction in the presence of the (hyper)nuclear 
medium is constructed,  based on the leading (irreducible) three-baryon forces derived within SU(3) chiral effective field theory.
We evaluate the contributions from three classes: contact terms, one-pion exchange and two-pion exchange.
In the strangeness-zero sector we recover the known result for the in-medium nucleon-nucleon interaction.
Explicit expressions for the $\Lambda N$ in-medium potential in (asymmetric) nuclear matter are presented.
Our results are suitable for implementation into calculations of (hyper)nuclear matter.
In order to estimate the low-energy constants of the leading three-baryon forces we introduce the decuplet baryons as 
explicit degrees of freedom and construct the relevant terms in the minimal non-relativistic Lagrangian.
With these, the constants are estimated through decuplet saturation.
Utilizing this approximation we provide numerical results for the effect of the three-body force in symmetric 
nuclear matter and pure neutron matter on the $\Lambda N$ interaction.
A moderate repulsion that increases with density is found in comparison to the free $\Lambda N$ interaction.
\end{abstract}



\maketitle
%\tableofcontents

\section{Introduction}

Three-body forces (3BFs) are an indispensable ingredient of any modern calculation of
few-nucleon systems. Specifically, for the three- and four-nucleon systems where
rigorous computations can be performed based on the Faddeev or Faddeev-Yakubovsky
equations there is clear evidence that agreement with experimental data cannot be
achieved if one resorts to nucleon-nucleon ($NN$) forces alone. 
Three-nucleon forces are required to 
reproduce correctly the binding energies in the few-nucleon sector but also
for scattering observables such as the proton-deuteron differential cross section
at incident proton energies around 100--200~MeV. 
For a recent review on these topics see, for example, \ct{Kalantar-Nayestanaki2012a}. 
Accordingly, one expects that such three-body forces are also important for
heavier nuclei as well as for the properties of nuclear matter. Indeed, in the
latter case standard calculations based on two-body interactions and
utilizing the Bethe-Goldstone equation are unable to describe the 
saturation point correctly, \ie to obtain the empirical energy per nucleon,
of $E/A = -16$~MeV, at the saturation density, $\rho_0 = 0.17$~fm$^{-3}$. 
Three-nucleon forces are considered as an essential mechanism that could resolve
this problem \ct*{Wiringa1988,Coraggio2014a, *[{}] [{, and references therein.}] Bogner2005, Krewald2012}. 

Likewise, three-body forces are expected also to play an important role in strangeness nuclear physics \ct*{Gal2016}, 
in particular the Lambda-nucleon-nucleon (\(\Lambda NN\)) interaction. 
It has been argued in the context of (exotic) neutron star matter 
that strongly repulsive 3BFs are needed in order to explain the recent 
observation of two-solar-mass neutron stars,
\ie to resolve the so-called hyperon puzzle \ct*{Takatsuka2008,Vidana2010,Lonardoni2013c,Lonardoni2013,Lonardoni2015}.
For example, a phenomenological $\Lambda NN$ three-body force has been introduced in Ref.\,\cite{Lonardoni2015}, with a repulsive coupling strength chosen large enough just so that the $\Lambda$ is prevented from appearing in dense matter and the equation-of-state remains sufficiently stiff to support a $2\,M_\odot$ neutron star.
The situation is less clear when it comes to light hypernuclei such as
the hypertriton $^3_\Lambda \rm H$, or $^4_\Lambda \rm H$ and $^4_\Lambda \rm He$,
owing to the fact that the two-body interaction in the relevant $\Lambda N$ and $\Sigma N$ 
systems is not well determined from the scarce experimental data presently available. 

Utilizing realistic models of the three-baryon force directly in many-body calculations
or in the Brueckner-Bethe-Goldstone approach (\eg via the Bethe-Faddeev 
equations \ct*{Bethe1965}) is a very challenging technical
task. Therefore, it has become customary to follow an alternative and simpler approach 
that consists in employing a density-dependent two-body interaction derived from the underlying three-body forces.
For the nucleonic sector such a density-dependent 
in-medium $NN$ interaction, generated at one-loop order by the leading chiral three-nucleon force,
has been constructed in \ct{Holt:2009ty}.
It has been shown in subsequent studies \ct*{Kaiser2012,Wellenhofer2014}  
and by several other calculations in the literature 
\ct*{Hebeler2010,Kohno2013a,Carbone:2014,Sammarruca2015,Kohno2015,Dyhdalo2016}
that his approximate treatment of three-body forces works very well. 

\begin{figure}[b]
$\text{NNLO:} \qquad \vcenter{\hbox{
\includegraphics[scale=0.5]{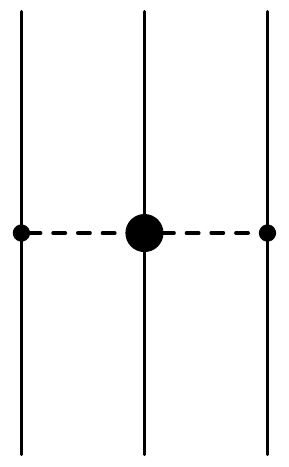}\qquad
\includegraphics[scale=0.5]{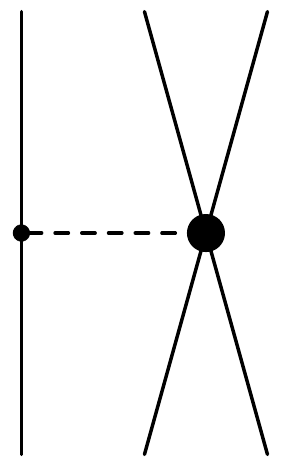}\qquad
\includegraphics[scale=0.5]{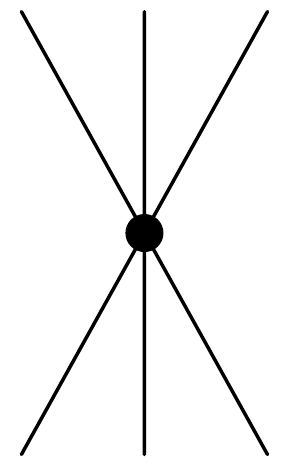}}}$
\caption{
Leading chiral three-baryon interactions: two-meson exchange, one-meson exchange and contact term.
\label{fig:3BF}
}
\end{figure}

In the present work we investigate the effect of the \(\Lambda NN\) three-body force on 
the \(\Lambda N\) interaction in the presence of a nuclear medium.
We start from the leading (irreducible) 3BFs, cf.\ \fig{fig:3BF}, 
which have been derived recently \ct*{Petschauer2016} within SU(3) chiral effective field theory (\cheft{}),
a systematic approach that exploits the symmetries of the underlying QCD.
Among other advantages, this approach ensures that the three-body forces are constructed 
consistently with the corresponding two-baryon interactions (e.g.\ $\Lambda N$, $\Sigma N$) \ct*{Polinder2006,Haidenbauer2013a}.
In our derivation we follow closely the work of \ct{Holt:2009ty} and
extend those calculations to sectors with non-zero strangeness.
As a result one obtains a density-dependent effective baryon-baryon interaction which facilitates the 
inclusion of effects from 3BFs into many-body calculations. 

The irreducible chiral 3BFs appear formally at next-to-next-to-leading order (NNLO). 
However, in the nucleonic sector one has observed that some of the corresponding low-energy constants (LECs) 
are much larger than expected from the hierarchy of nuclear forces. This feature has its physical origin in 
the strong coupling of the \(\pi N\) system to the low-lying \(\Delta(1232)\)-resonance.
It is therefore natural to include the \(\Delta(1232)\)-isobar as an explicit degree of freedom 
in the chiral Lagrangian (\cf \cts{Jenkins1991b,VanKolck1994,Hemmert1997,Kaiser1998,Krebs2007}). %Bernard1997
The small mass difference between nucleons and deltas (\(293\ \mathrm{MeV}\)) introduces a small scale, which 
can be included consistently in the chiral power counting scheme and the hierarchy of nuclear forces.
The dominant part of the three-nucleon interaction mediated by two-pion exchange and virtual \(\Delta(1232)\) excitation is 
then promoted to next-to-leading order (NLO). The appearance of 
the inverse mass splitting explains the large numerical values of the corresponding 
LECs \ct*{Bernard1997,Epelbaum2002a,Epelbaum2008a,Epelbaum2009}.

\begin{figure}
$\text{NLO:} \qquad \vcenter{\hbox{
\includegraphics[scale=.5]{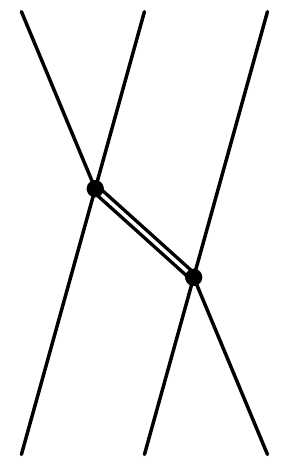}\qquad
\includegraphics[scale=.5]{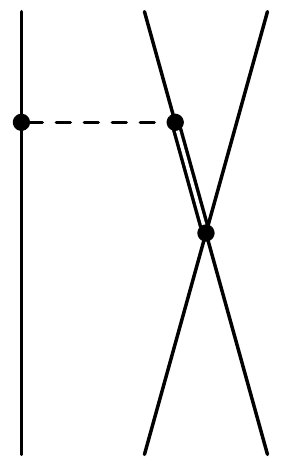}\qquad
\includegraphics[scale=.5]{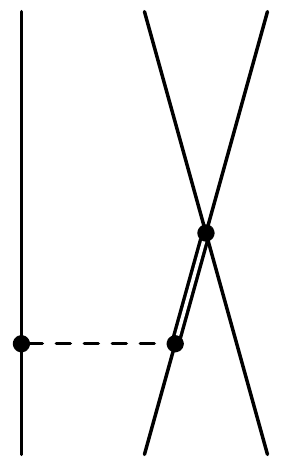}\qquad
\includegraphics[scale=.5]{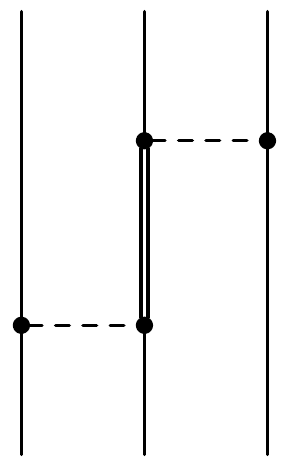}}}$
\caption{
Three-baryon forces arising from virtual decuplet excitation (represented by double lines).
\label{fig:hierdec}
}
\end{figure}

In SU(3) \cheft{} the situation is similar. Specifically, 
in systems with strangeness \(S=-1\), like \(\Lambda NN\), intermediate baryons such as the spin-3/2 \(\Sigma^*\)(1385)-resonance could 
play an analogous role as the \(\Delta(1232)\) in the \(NNN\) system.
Indeed the decuplet-octet mass splittings are on average smaller than the delta-nucleon splitting.
Also in SU(3) \cheft{}
the mass splitting (in the chiral limit) should be counted together with external momenta and meson masses 
as \(\mathcal O(q)\) and therefore parts of the NNLO three-baryon interaction are promoted to NLO by the 
explicit inclusion of the baryon decuplet, as illustrated in \fig{fig:hierdec} (see also \cts{VanKolck1994,Meissner2008,Epelbaum2008a}). 
One expects that these NLO contributions give the dominant part of the 3BFs and thus should provide a reasonable basis for investigating the effects of the \(\Lambda NN\) 
interaction. Of particular interest is the long-range contribution arising from two-pion 
exchange.

In the present paper we exploit the mechanism of decuplet saturation to estimate the strengths of chiral 3BFs.
By including decuplet baryons not only parts of the two-pion exchange 3BF are promoted to NLO but also contributions that involve contact vertices.
This is illustrated in \fig{fig:hierdec}.
In the purely nucleonic case such contributions do not arise because
a leading-order \(\Delta NNN\) four-baryon contact vertex is forbidden by the Pauli principle.
The decuplet induced 3BF of short range still involve two unknown parameters and, therefore, a reliable quantitative estimate of 3BF effects in the strangeness \(S=-1\) sector is difficult to make at present.  
Contrary to the practice in the nucleonic sector, 
a direct determination of the LECs from 
experimental information on few-baryon systems with strangeness \(S=-1\) is not (yet) feasible
because of the limited amount and accuracy of the data.

This paper is organized as follows.
In \sect{sec:med} we present the general expressions for the effective two-baryon potential derived from the irreducible chiral three-baryon forces for all strangeness sectors.
As an example we give the explicit results for the \(\Lambda N\) interaction in symmetric and asymmetric nuclear matter.
In \sect{sec:dec} we introduce the pertinent chiral Lagrangians including decuplet baryons and estimate the LECs of the 3BFs via decuplet saturation.
Finally, in \sect{sec:res}, we present numerical results for the in-medium \(\Lambda N\) interaction within this approximation.
In the appendices we collect for comparison the explicit expressions for the antisymmetrized \(NN\) in-medium interaction in isospin-symmetric nuclear matter.
Furthermore, details related to the construction of the decuplet Lagrangian are presented.

In this work we consider only those medium corrections which arise from irreducible three-baryon forces.
Further density-dependent contributions originating from reducible three-baryon processes are also known to be important.
A prominent example is the reducible \(\Lambda NN\) interaction involving two-pion exchange and a \(\Sigma\) hyperon in the intermediate state.
An investigation of these reducible contributions in the many-body sector goes beyond the scope of the present paper.
In proper few-body calculations incorporating both $\Lambda$ and $\Sigma$ hyperons as explicit degrees of freedom, such reducible contributions are generated automatically by coupled-channel Faddeev and Yakubovsky equations from iterated $\Lambda N\leftrightarrow\Sigma N$ interactions \ct*{Nogga2014a}.

\section{In-medium baryon-baryon interaction} \label{sec:med}

\begin{figure}
\begin{minipage}{.10\textwidth} \centering
\includegraphics[scale=.6]{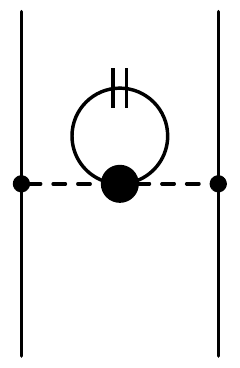} \\ (1)
\end{minipage} \quad
\begin{minipage}{.12\textwidth} \centering
\includegraphics[scale=.6]{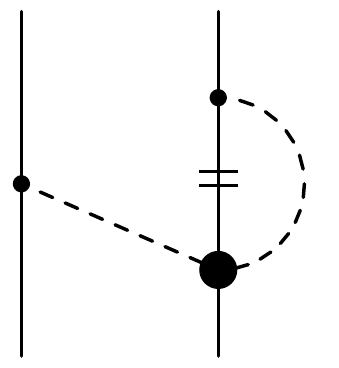} \\ (2a)
\end{minipage} \
\begin{minipage}{.12\textwidth} \centering
\includegraphics[scale=.6]{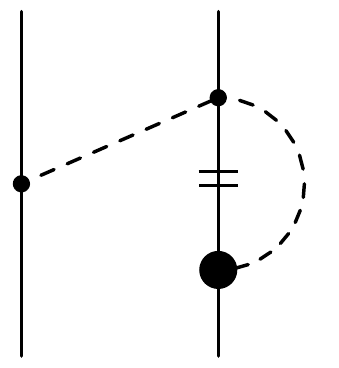} \\ (2b)
\end{minipage} \
\begin{minipage}{.10\textwidth} \centering
\includegraphics[scale=.6]{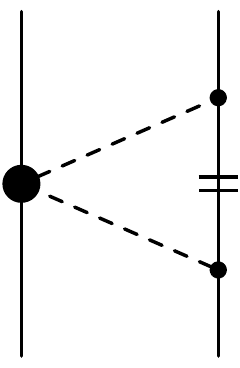} \\ (3)
\end{minipage} \quad
\begin{minipage}{.12\textwidth} \centering
\includegraphics[scale=.6]{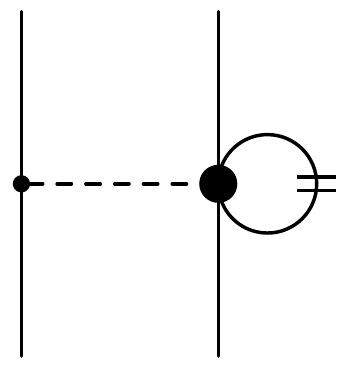} \\ (4)
\end{minipage} \
\begin{minipage}{.10\textwidth} \centering
\includegraphics[scale=.6]{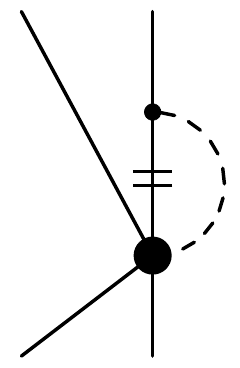} \\ (5a)
\end{minipage} \
\begin{minipage}{.10\textwidth} \centering
\includegraphics[scale=.6]{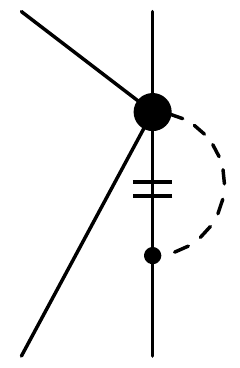} \\ (5b)
\end{minipage} \
\begin{minipage}{.10\textwidth} \centering
\includegraphics[scale=.6]{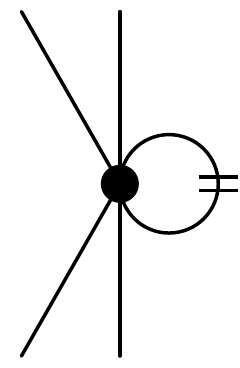} \\ (6)
\end{minipage}
\caption{
Effective two-baryon interaction from genuine three-baryon forces. Contributions arise from 
two-pion exchange (1), (2a), (2b), (3), one-pion exchange (4), (5a), (5b) and the contact interaction (6).
\label{fig:med}
}
\end{figure}

In this section we derive the effect of a three-body force on the baryon-baryon interaction in the presence of 
a (hyper)nuclear medium.
We follow closely the work of \ct{Holt:2009ty}, where density-dependent corrections to the \(NN\) interaction have 
been calculated from leading-order chiral three-nucleon forces.
In order to obtain an effective baryon-baryon interaction from the irreducible 3BFs 
in \fig{fig:3BF}, one closes two baryon lines which represents diagrammatically the sum over occupied states within 
the Fermi sea.
Such a ``medium insertion'' is symbolized by short double lines on a baryon propagator.
All types of diagrams arising this way are shown in \fig{fig:med}.

We restrict ourselves to the contact term and to the contributions from one-  
and two-pion exchange processes which are expected to be dominant. Hence, the calculation is done for equal meson masses. 
In principle, within SU(3) \cheft{} further contributions arise that involve the exchange of at least one heavier meson 
(kaon or eta meson). At moderate densities these contributions of much shorter range can  
effectively be absorbed into a contact term representing the short-range part of the three-baryon force.
When evaluating diagrams the medium insertion provides the factor \(-2\pi\delta(k_0)\theta(k_f-|\vec k|)\). An additional minus sign comes from a closed fermion loop.
Equivalently, the effective two-body interaction can be constructed from the expressions for the three-baryon potentials in \ct{Petschauer2016} via the relation
\begin{equation} \label{eq:red}
V_{12} = \sum_B\tr_{\sigma_3}\int_{|\vec k|\leq k_f^B}\frac{\mathrm d^3k}{(2\pi)^3} V_{123} \,,
\end{equation}
where \(\tr_{\sigma_3}\) denotes the spin trace over the third particle and the sum goes over all baryon species \(B\) in the Fermi sea (with Fermi momentum \(k_f^B\)).
In the following, we derive the general expressions of the effective potentials for a single baryon species \(B\).
The full potential is given by a sum over all species.
The density of the baryon species \(B\) is given by
\begin{equation}
\rho_B = 2 \int_{|\vec k|\leq k_f^B}\frac{\mathrm d^3k}{(2\pi)^3} = \frac{{(k_f^B)}^3}{3\pi^2}\,,
\end{equation}
and the full density is obtained by summing over all species in the (hyper)nuclear medium, \(\rho = \sum_B \rho_B\).

As done in \ct{Holt:2009ty}, we consider the scattering of two baryons within the medium in the center-of-mass frame
\begin{equation}
B_1(\vec p\,) + B_2(-\vec p\,) \to B_3(\vec p^{\,\prime}) + B_4(-\vec p^{\,\prime}) \, ,
\end{equation}
for on-shell kinematics: \(p^2=p'^2\).
For direct diagrams the relevant momentum transfer is \(\vec q = \vec p^{\,\prime}-\vec p\), 
for the exchange-type diagrams the relevant momentum transfer is \(\vec k = \vec p^{\,\prime}+\vec p\).

In the course of the calculation one encounters integrals of one pion propagator or the product of two pion propagators over a Fermi sphere.
The loop functions \(\Gamma_i\) involving a single pion propagator are defined by
\begin{equation}
\int_{|\vec l|\leq k_f^B}\frac{\mathrm d^3l}{2\pi}
\frac{1}{\m^2+(\vec l+\vec p\,)^2}
\begin{pmatrix}
1 \\ \vec l \\ \vec l\otimes\vec l
\end{pmatrix}
= \begin{pmatrix}
\Gamma_0(p,k_f^B) \\ \vec p\ \Gamma_1(p,k_f^B) \\ \mathbbm1\Gamma_2(p,k_f^B)+\vec p\otimes\vec p\, \Gamma_3(p,k_f^B)
\end{pmatrix}.
\end{equation}
The loop functions \(G_i\) involving two different pion propagators are given by
\begin{align} \label{eq:Gs}
&\int_{|\vec l|\leq k_f^B}\frac{\mathrm d^3l}{2\pi} \frac{1}{[\m^2+(\vec l+\vec p\,)^2][\m^2+(\vec l+\vec p^{\,\prime})^2]}
\begin{pmatrix}
1 \\ \vec l \\ \vec l\otimes\vec l \\ \\ l^2 \\ l^2\vec l \\ l^4
\end{pmatrix}
= \begin{pmatrix}
G_0(p,q,k_f^B) \\ 
(\vec p^{\,\prime}+\vec p\,)G_1(p,q,k_f^B) \\ 
\mathbbm1 G_2(p,q,k_f^B)+(\vec p^{\,\prime}+\vec p\,)\otimes (\vec p^{\,\prime}+\vec p\,) G_3(p,q,k_f^B) \\ 
+(\vec p^{\,\prime}-\vec p\,)\otimes (\vec p^{\,\prime}-\vec p\,) G_4(p,q,k_f^B) \\ 
G_*(p,q,k_f^B) \\ 
(\vec p^{\,\prime}+\vec p\,)G_{1*}(p,q,k_f^B) \\ 
G_{**}(p,q,k_f^B)
\end{pmatrix} .
\end{align}
The explicit formulas for the loop functions can be found in Section III.A.\ of \ct{Holt:2009ty}.
Note that in some cases the expression on the left-hand side of \eq{eq:Gs} (with two pion propagators) appears with the substitution \(\vec p^{\,\prime}\to -\vec p^{\,\prime}\).
Consequently, this substitution has also to be done on the right-hand side and the arguments of \(G_i\) are changed to \(G_i(p,k,k_f^B)\).

\subsection{Contributions from two-pion exchange}

Let us start with the two-pion exchange contribution to the in-medium baryon-baryon interaction.
The corresponding three-baryon potential for a prototype two-meson exchange diagram is given in Eq.\ (34) in \ct{Petschauer2016} and it reads
\begin{equation} \label{eq:gen2mes}
V ={} -\frac{1}{4f_0^4} \frac{\vec\sigma_A\cdot\vec q_{li}\ \vec\sigma_C\cdot\vec q_{nk}}{(\vec q_{li}^{\,2}+m_{\phi_1}^2)(\vec q_{nk}^{\,2}+m_{\phi_2}^2)}
\Big(N_{\substack{lmn\\ijk}}^1 + N_{\substack{lmn\\ijk}}^2\,\vec q_{li}\cdot\vec q_{nk} +N_{\substack{lmn\\ijk}}^3\,\mathrm i\,(\vec q_{li}\times\vec q_{nk})\cdot\vec\sigma_B\Big) \,,
\end{equation}
adopting the same definitions and conventions as in \ct{Petschauer2016}.
The quantity \(f_0=93.0~\mathrm{MeV}\) is the meson-decay constant in the chiral limit and \(m_{\phi_1},m_{\phi_2}\) are the masses of the two exchanged meson. The potential \(V\) in \eq{eq:gen2mes}
involves a variety of combinations of SU(3) factors 
\begin{align*}
N^1_{\substack{lmn\\ijk}} ={}& N_{B_lB_i\bar\phi_1}N_{B_nB_k\phi_2}   \sum_{c^f=b_D,b_F,b_0}\frac{c^f}{4} ( N^f_{\phi_1\substack{m\\j}\bar\phi_2} + N^f_{\bar\phi_2\substack{m\\j}\phi_1}) \,, \\
N^2_{\substack{lmn\\ijk}} ={}& -N_{B_lB_i\bar\phi_1}N_{B_nB_k\phi_2}   \sum_{c^f=b_1,b_2,b_3,b_4}c^f ( N^f_{\phi_1\substack{m\\j}\bar\phi_2} + N^f_{\bar\phi_2\substack{m\\j}\phi_1}) \,, \\
N^3_{\substack{lmn\\ijk}} ={}& N_{B_lB_i\bar\phi_1}N_{B_nB_k\phi_2}   \sum_{c^f=d_1,d_2,d_3}c^f ( N^f_{\phi_1\substack{m\\j}\bar\phi_2} - N^f_{\bar\phi_2\substack{m\\j}\phi_1}) \,. \numberthis
\end{align*}
Following the detailed exposition of all possible diagrams in Fig.\ 5 of \ct{Petschauer2016}, we close the two baryon lines \(B_3\) and \(B_6\) for each three-body diagram.
This leads to the topologies (1), (2a), (2b), and (3) shown in \fig{fig:med}.

Performing the spin trace and integrating over a Fermi sphere according to \eq{eq:red} for the diagrams leading to the topology (1) in \fig{fig:med}, one obtains for the direct effective potential (involving the momentum transfer \(\vec q = \vec p^{\,\prime} - \vec p\,\)) generated by the Fermi sea of the baryon species \(B\)
\begin{equation} \label{med1d}
V^\mathrm{med,1,D} = \frac{\rho_B}{4f_0^4}\frac{\vec\sigma_1\cdot\vec q\ \vec\sigma_2\cdot\vec q}{(q^2+\m^2)^2} \, ( N^1_{\substack{4B3\\2B1}} - N^2_{\substack{4B3\\2B1}} q^2 ) \,.
\end{equation}
The resulting interaction is proportional to the density \(\rho_B\) of the baryon species \(B\).
The exchange diagrams (involving the momentum transfer \(\vec k = \vec p^{\,\prime} + \vec p\,\)) lead to the same contribution with different factors \(N\):
\begin{equation} \label{med1e}
V^\mathrm{med,1,E} = -P^{(\sigma)}\frac{\rho_B}{4f_0^4}\frac{\vec\sigma_1\cdot\vec k\ \vec\sigma_2\cdot\vec k}{(k^2+\m^2)^2} \, ( N^1_{\substack{3B4\\2B1}} - N^2_{\substack{3B4\\2B1}} k^2 ) \, .
\end{equation}
It can be obtained from the direct contribution by multiplying with the negative spin-exchange operator, \(-P^{(\sigma)} = -\frac12(\mathbbm1+\vec\sigma_1\cdot\vec\sigma_2)\), and substituting \(\vec p^{\,\prime}\to-\vec p^{\,\prime}\) (and therefore also \(\vec q\to-\vec k\)).

Similarly, the topology (2a) of \fig{fig:med} gives rise to the following direct potential: 
\begin{align*} \label{med2ad}
V^\mathrm{med,2a,D} ={}& -\frac1{16\pi^2f_0^4(q^2+\m^2)}\Bigg\{
X_1 \vec\sigma_1\cdot\vec q\ \vec\sigma_2\cdot\vec q\   \big[ \frac23{(k_f^B)}^3-\m^2\Gamma_0(p,k_f^B) \big] \\
& + \big(X_2 \vec\sigma_1\cdot\vec q\ \vec\sigma_2\cdot\vec p^{\,\prime}+ X_3 \vec\sigma_1\cdot\vec p^{\,\prime}\ \vec\sigma_2\cdot\vec q\, \big)  \big[ \Gamma_0(p,k_f^B)+\Gamma_1(p,k_f^B) \big] \\
& + \big(X_4 \vec\sigma_1\cdot\vec q\ \vec\sigma_2\cdot\vec p^{\,\prime} + X_5\vec\sigma_1\cdot\vec p^{\,\prime}\ \vec\sigma_2\cdot\vec q\,\big)  \frac{q^2}2 \big[\Gamma_0(p,k_f^B)+2\Gamma_1(p,k_f^B)+\Gamma_3(p,k_f^B)\big] \\
& + (X_4+X_5) \vec\sigma_1\cdot\vec q\ \vec\sigma_2\cdot\vec q\ \Gamma_2(p,k_f^B) \numberthis
\Bigg\} \,,
\end{align*}
written in terms of the loop functions \(\Gamma_i\).
The relevant SU(3) factors \(X_i\) are given in the first column of \tab{tab:med2}.
%%%
\begin{table}
\centering
\renewcommand*{\arraystretch}{1.4}
\begin{tabular}{>{\(}c<{\)}>{\(}c<{\)}>{\(}c<{\)}>{\(}c<{\)}>{\(}c<{\)}}
\toprule
X & \text{med,2a,D} & \text{med,2a,E} & \text{med,2b,D} & \text{med,2b,E} \\
\cmidrule(lr){1-1} \cmidrule(lr){2-5}
X_1 & N^3_{\substack{3B4\\12B}} - N^3_{\substack{3B4\\B12}} & N^3_{\substack{4B3\\12B}} - N^3_{\substack{4B3\\B12}} & N^3_{\substack{B43\\2B1}} - N^3_{\substack{43B\\2B1}} & N^3_{\substack{B34\\2B1}} - N^3_{\substack{34B\\2B1}} \\
X_2 & N^1_{\substack{3B4\\12B}} & N^1_{\substack{4B3\\12B}} & N^1_{\substack{B43\\2B1}} & N^1_{\substack{B34\\2B1}} \\
X_3 & N^1_{\substack{3B4\\B12}} & N^1_{\substack{4B3\\B12}} & N^1_{\substack{43B\\2B1}} & N^1_{\substack{34B\\2B1}} \\
X_4 & - N^2_{\substack{3B4\\12B}} - N^3_{\substack{3B4\\12B}} & - N^2_{\substack{4B3\\12B}} - N^3_{\substack{4B3\\12B}} & - N^2_{\substack{B43\\2B1}} - N^3_{\substack{B43\\2B1}} & - N^2_{\substack{B34\\2B1}} - N^3_{\substack{B34\\2B1}} \\
X_5 & N^3_{\substack{3B4\\B12}} - N^2_{\substack{3B4\\B12}} & N^3_{\substack{4B3\\B12}} - N^2_{\substack{4B3\\B12}} & N^3_{\substack{43B\\2B1}} - N^2_{\substack{43B\\2B1}} & N^3_{\substack{34B\\2B1}} - N^2_{\substack{34B\\2B1}} \\
\bottomrule
\end{tabular}
\renewcommand*{\arraystretch}{1.0}
\caption{SU(3) factors for the two-meson-exchange contributions of type (2).}
\label{tab:med2}
\end{table}
%%%
The exchange diagrams lead to the same contributions as the direct ones, again multiplying with \(-P^{(\sigma)}\) and making the substitution \(\vec p^{\,\prime}\to-\vec p^{\,\prime}\) (and therefore also \(\vec q\to-\vec k\)):
\begin{equation} \label{med2ae}
V^\mathrm{med,2a,E} = -P^{(\sigma)} V^\mathrm{med,2a,D}\Big|_{\vec p^{\,\prime}\to-\vec p^{\,\prime}} \,.
\end{equation} 
Furthermore, the coefficients \(X_i\) have to be inserted according to the second column of \tab{tab:med2}.
The reflected topology (2b) in \fig{fig:med} leads to a result similar to that of topology (2a):
\begin{align*} \label{med2bd}
V^\mathrm{med,2b,D} ={}& -\frac1{16\pi^2f_0^4(q^2+\m^2)}\Bigg\{
X_1 \vec\sigma_1\cdot\vec q\ \vec\sigma_2\cdot\vec q\   \big[ \frac23{(k_f^B)}^3-\m^2\Gamma_0(p,k_f^B) \big] \\
& - \big(X_2 \vec\sigma_1\cdot\vec q\ \vec\sigma_2\cdot\vec p+ X_3 \vec\sigma_1\cdot\vec p \ \vec\sigma_2\cdot\vec q\, \big)  \big[ \Gamma_0(p,k_f^B)+\Gamma_1(p,k_f^B) \big] \\
& - \big(X_4 \vec\sigma_1\cdot\vec q\ \vec\sigma_2\cdot\vec p  + X_5\vec\sigma_1\cdot\vec p \ \vec\sigma_2\cdot\vec q\,\big)  \frac{q^2}2 \big[\Gamma_0(p,k_f^B)+2\Gamma_1(p,k_f^B)+\Gamma_3(p,k_f^B)\big] \\
& + (X_4+X_5) \vec\sigma_1\cdot\vec q\ \vec\sigma_2\cdot\vec q\ \Gamma_2(p,k_f^B)
\Bigg\} \,, \numberthis
\end{align*}
where the SU(3) factors \(X_i\) are now given in the third column of \tab{tab:med2}.
For the corresponding exchange diagrams one obtains again
\begin{equation} \label{med2be}
V^\mathrm{med,2b,E} = -P^{(\sigma)} V^\mathrm{med,2b,D}\Big|_{\vec p^{\,\prime}\to-\vec p^{\,\prime}} \,,
\end{equation} 
with the \(X_i\) listed in the fourth column of \tab{tab:med2}.

The diagrams contributing to the topology (3) in \fig{fig:med} lead to the following direct potential
\begin{align*} \label{med3d}
V^\mathrm{med,3,D} ={}& -\frac1{16\pi^2f_0^4}\Bigg\{
X_1 \frac12\big[2\Gamma_0(p,k_f^B)-(q^2+2\m^2)G_0(p,q,k_f^B)\big] \\ &
+ X_2 \frac14\big[ \frac83 {(k_f^B)}^3 - 4(q^2+2\m^2)\Gamma_0(p,k_f^B) - 2q^2 \Gamma_1(p,k_f^B) + (q^2+2\m^2)^2G_0(p,q,k_f^B) \big] \\ &
+ X_3 \big[G_0(p,q,k_f^B)+4G_1(p,q,k_f^B)+4G_3(p,q,k_f^B)\big] (\vec q\times\vec p\,)\cdot\vec\sigma_1\, (\vec q\times\vec p\,)\cdot\vec\sigma_2 \\ &
+ X_3 G_2(p,q,k_f^B) (q^2\vec\sigma_1\cdot\vec\sigma_2-\vec\sigma_1\cdot\vec q\ \vec\sigma_2\cdot\vec q\,) \\ &
+ \frac{\mathrm i}2 (\vec q\times\vec p\,)\cdot(\vec\sigma_1+\vec\sigma_2) \Big[ X_4 \Big(G_0(p,q,k_f^B)+2G_1(p,q,k_f^B) \Big) \\
&\qquad + X_5 \frac12 \Big(2\Gamma_0(p,k_f^B)+2\Gamma_1(p,k_f^B)-(q^2+2\m^2)\big(G_0(p,q,k_f^B)+2G_1(p,q,k_f^B)\big)\Big) \Big] \\ &
+ \frac{\mathrm i}2 (\vec q\times\vec p\,)\cdot(\vec\sigma_1-\vec\sigma_2) \Big[ X_6\Big(G_0(p,q,k_f^B)+2G_1(p,q,k_f^B)\Big) \\
&\qquad + X_7 \frac12 \Big(2\Gamma_0(p,k_f^B)+2\Gamma_1(p,k_f^B)-(q^2+2\m^2)\big(G_0(p,q,k_f^B)+2G_1(p,q,k_f^B)\big)\Big) \Big]
\Bigg\} \,, \numberthis
\end{align*}
where the new SU(3) factors \(X_i\) are given in the first column of \tab{tab:med3}.
In order to write out this potential both loop functions \(\Gamma_i\) and \(G_i\) are needed.
%%%
\begin{table}
\centering
\renewcommand*{\arraystretch}{1.4}
\begin{tabular}{>{\(}c<{\)}>{\(}c<{\)}>{\(}c<{\)}}
\toprule
X & \text{med,3,D} & \text{med,3,E} \\
\cmidrule(lr){1-1} \cmidrule(lr){2-3}
X_1 & N^1_{\substack{43B\\B12}} + N^1_{\substack{B43\\12B}} & N^1_{\substack{34B\\B12}} + N^1_{\substack{B34\\12B}} \\
X_2 & - N^2_{\substack{43B\\B12}} - N^2_{\substack{B43\\12B}} & - N^2_{\substack{34B\\B12}} - N^2_{\substack{B34\\12B}} \\
X_3 & N^3_{\substack{43B\\B12}} - N^3_{\substack{B43\\12B}} & N^3_{\substack{34B\\B12}} - N^3_{\substack{B34\\12B}} \\
X_4 & N^1_{\substack{43B\\B12}} + N^1_{\substack{B43\\12B}} & N^1_{\substack{34B\\B12}} + N^1_{\substack{B34\\12B}} \\
X_5 & -N^2_{\substack{43B\\B12}}-N^3_{\substack{43B\\B12}} - N^2_{\substack{B43\\12B}} + N^3_{\substack{B43\\12B}} & -N^2_{\substack{34B\\B12}}-N^3_{\substack{34B\\B12}} - N^2_{\substack{B34\\12B}} + N^3_{\substack{B34\\12B}} \\
X_6 & - N^1_{\substack{43B\\B12}} + N^1_{\substack{B43\\12B}} & - N^1_{\substack{34B\\B12}} + N^1_{\substack{B34\\12B}} \\
X_7 & N^2_{\substack{43B\\B12}} - N^3_{\substack{43B\\B12}} - N^2_{\substack{B43\\12B}} - N^3_{\substack{B43\\12B}} & N^2_{\substack{34B\\B12}} - N^3_{\substack{34B\\B12}} - N^2_{\substack{B34\\12B}} - N^3_{\substack{B34\\12B}} \\
\bottomrule
\end{tabular}
\renewcommand*{\arraystretch}{1.0}
\caption{SU(3) factors for two-meson-exchange contribution of type (3).}
\label{tab:med3}
\end{table}
%%%
The exchange diagrams lead to the same contribution, multiplying with \(-P^{(\sigma)}\) and substituting \(\vec q\to-\vec k\):
\begin{equation} \label{med3e}
V^\mathrm{med,3,E} = -P^{(\sigma)} V^\mathrm{med,3,D}\Big|_{\vec p^{\,\prime}\to-\vec p^{\,\prime}} \,,
\end{equation} 
where the appropriate combinations \(X_i\) are given in the second column of \tab{tab:med3}.

\subsection{Contributions from one-pion exchange}

Let us now turn to the one-meson exchange three-baryon interaction.
We take the prototype one-meson exchange potentials (written in Eq.\ (29) of \ct{Petschauer2016}) and antisymmetrize the four-baryon contact vertex, (this means the four diagrams in each line of Fig.\ 3 of \ct{Petschauer2016} are summed up).
This leads to the expression:
\begin{equation}
V = \frac{1}{2f_0^2} \frac{\vec\sigma_A\cdot\vec q_{li}}{\vec q_{li}^{\,2}+m_{\phi}^2} \Big(
N^1_{\substack{lmn\\ijk}} \vec\sigma_B\cdot\vec q_{li}
+N^2_{\substack{lmn\\ijk}} \vec\sigma_C\cdot\vec q_{li}
+N^3_{\substack{lmn\\ijk}} \mathrm i\,(\vec\sigma_B\times\vec\sigma_C)\cdot\vec q_{li}
\Big)\,,
\end{equation}
where the momentum transfer \(\vec q_{li}\) is given by \(\vec q_{li} = \vec p_l-\vec p_i\) and the new SU(3) coefficients read
\begin{align*}
N^1_{\substack{lmn\\ijk}} &= N_{B_lB_i\phi}\Big(
\sum_{f=1}^{10} D_f N^{f}_{\substack{nm\\kj}\,\bar\phi}
- \tfrac12\sum_{f=1}^{10} D_f N^{f}_{\substack{nm\\jk}\,\bar\phi}
- \sum_{f=11}^{14} D_f N^{f}_{\substack{nm\\jk}\,\bar\phi}
- \tfrac12\sum_{f=1}^{10} D_f N^{f}_{\substack{mn\\kj}\,\bar\phi}
+ \sum_{f=11}^{14} D_f N^{f}_{\substack{mn\\kj}\,\bar\phi}
\Big) \,, \\
N^2_{\substack{lmn\\ijk}} &= N_{B_lB_i\phi}\Big(
\sum_{f=1}^{10} D_f N^{f}_{\substack{mn\\jk}\,\bar\phi}
- \tfrac12\sum_{f=1}^{10} D_f N^{f}_{\substack{nm\\jk}\,\bar\phi}
+ \sum_{f=11}^{14} D_f N^{f}_{\substack{nm\\jk}\,\bar\phi}
- \tfrac12\sum_{f=1}^{10} D_f N^{f}_{\substack{mn\\kj}\,\bar\phi}
- \sum_{f=11}^{14} D_f N^{f}_{\substack{mn\\kj}\,\bar\phi}
\Big) \,, \\
N^3_{\substack{lmn\\ijk}} &= N_{B_lB_i\phi}\Big(
\sum_{f=11}^{14} D_f N^{f}_{\substack{mn\\jk}\,\bar\phi}
- \sum_{f=11}^{14} D_f N^{f}_{\substack{nm\\kj}\,\bar\phi}
+ \frac12\sum_{f=1}^{10} D_f N^{f}_{\substack{nm\\jk}\,\bar\phi}
- \frac12\sum_{f=1}^{10} D_f N^{f}_{\substack{mn\\kj}\,\bar\phi}
\Big)\Big] \,. \numberthis
\end{align*}
Next we have to consider all 9 rows of diagrams in Fig.\ 3 of \ct{Petschauer2016} corresponding to all possibilities to close the baryon lines \(B_3\) and \(B_6\).
This procedure leads to the three one-pion exchange topologies (4), (5a) and (5b) displayed \fig{fig:med}.

Topology (4) gives rise to a direct contribution to the in-medium baryon-baryon potential of the form
\begin{equation} \label{med4d}
V^\mathrm{med,4,D} = \frac{\rho_B}{2f_0^2(q^2+\m^2)}\vec\sigma_1\cdot\vec q\ \vec\sigma_2\cdot\vec q\,( N^1_{\substack{34B\\12B}} + N^2_{\substack{4B3\\2B1}} ) \,,
\end{equation}
which depends linearly on the density \(\rho_B\) of baryon species \(B\).
Similarly, the exchange diagrams yield 
\begin{equation} \label{med4e}
V^\mathrm{med,4,E} = \frac{\rho_B}{4f_0^2(k^2+\m^2)}P^{(\sigma)}\vec\sigma_1\cdot\vec k\ \vec\sigma_2\cdot\vec k\,(
N^1_{\substack{4B3\\12B}}+N^2_{\substack{4B3\\12B}}+2N^3_{\substack{4B3\\12B}} +
N^1_{\substack{34B\\2B1}}+N^2_{\substack{34B\\2B1}}-2N^3_{\substack{34B\\2B1}} ) \,.
\end{equation}

Furthermore, one obtains from topology (5a) in \fig{fig:med} the following effective baryon-baryon potential:
\begin{align*} \label{med5a}
V^{\mathrm{med},5a} = \frac{1}{8\pi^2f_0^2} \bigg\{ &
(X_1+X_2\vec\sigma_1\cdot\vec\sigma_2)\big(\frac23{(k_f^B)}^3-\m^2\Gamma_0(p,k_f^B)\big) \\ &
+ X_3 \big[ \vec\sigma_1\cdot\vec p^{\,\prime} \ \vec\sigma_2\cdot\vec p^{\,\prime} \big(\Gamma_0(p,k_f^B)+2\Gamma_1(p,k_f^B)+\Gamma_3(p,k_f^B)\big) + \Gamma_2(p,k_f^B)\vec\sigma_1\cdot\vec\sigma_2 \big]
\bigg\} \,, \numberthis
\end{align*}
where the coefficients \(X_i\) are listed in the first column of \tab{tab:med5}.
%%%
\begin{table}
\centering
\renewcommand*{\arraystretch}{1.4}
\begin{tabular}{>{\(}c<{\)}>{\(}c<{\)}>{\(}c<{\)}}
\toprule
X & \text{med,5a} & \text{med,5b} \\
\cmidrule(lr){1-1} \cmidrule(lr){2-3}
X_1 & \frac12\big(
N^1_{\substack{34B\\B12}} +N^2_{\substack{34B\\B12}} + 2N^3_{\substack{34B\\B12}}
+N^1_{\substack{4B3\\B12}} +N^2_{\substack{4B3\\B12}} - 2N^3_{\substack{4B3\\B12}}
\big ) &
\frac12\big(
N^1_{\substack{B34\\2B1}} +N^2_{\substack{B34\\2B1}} + 2N^3_{\substack{B34\\2B1}}
+N^1_{\substack{B34\\12B}} +N^2_{\substack{B34\\12B}} - 2N^3_{\substack{B34\\12B}}
\big ) \\
X_2 & \frac12\big(
N^2_{\substack{34B\\B12}} -N^1_{\substack{34B\\B12}}
+N^1_{\substack{4B3\\B12}} -N^2_{\substack{4B3\\B12}}
\big ) &
\frac12\big(
N^1_{\substack{B34\\2B1}} -N^2_{\substack{B34\\2B1}}
+N^2_{\substack{B34\\12B}} -N^1_{\substack{B34\\12B}}
\big ) \\
X_3 & 
N^1_{\substack{34B\\B12}} -N^3_{\substack{34B\\B12}}
+N^2_{\substack{4B3\\B12}} +N^3_{\substack{4B3\\B12}} &
N^2_{\substack{B34\\2B1}} -N^3_{\substack{B34\\2B1}}
+N^1_{\substack{B34\\12B}} +N^3_{\substack{B34\\12B}} \\
\bottomrule
\end{tabular}
\renewcommand*{\arraystretch}{1.0}
\caption{SU(3) factors for one-meson-exchange contributions of type (5).}
\label{tab:med5}
\end{table}
%%%
The mirror topology (5b) in \fig{fig:med} gives rise to the same result, but with the replacement \(\vec p^{\,\prime} \to \vec p\), due to the reflection of the pion loop from the final state into the initial state:
\begin{equation} \label{med5b}
V^{\mathrm{med},5b}=V^{\mathrm{med},5a}\vert_{\vec p^{\,\prime} \to \vec p} \,,
\end{equation} 
and with the modified coefficients \(X_i\) given in the second column of \tab{tab:med5}.

\subsection{Contributions from contact terms}

Finally, we have to study the contact interaction, this means the topology (6) in \fig{fig:med}.
The fully antisymmetrized three-baryon contact potential reads (see Eqs.~(8) and (12) in \ct*{Petschauer2016})
\begin{equation}
V = -\Big[N^1_{\substack{456\\123}}+N^2_{\substack{456\\123}}\vec\sigma_1\cdot\vec\sigma_2+N^3_{\substack{456\\123}}\vec\sigma_1\cdot\vec\sigma_3+N^4_{\substack{456\\123}}\vec\sigma_2\cdot\vec\sigma_3+N^5_{\substack{456\\123}}\,\mathrm i\, \vec\sigma_1\cdot(\vec\sigma_2\times\vec\sigma_3)\Big] \,.
\end{equation}
Here, the spin trace over the third particle eliminates the last three term and the (trivial) Fermi sphere integration gives a factor \(\rho_B\), such that one obtains the following momentum-independent in-medium potential:
\begin{equation} \label{med6}
V^{\mathrm{med},6} = -\rho_B(N^1_{\substack{34B\\12B}}+N^2_{\substack{34B\\12B}}\vec\sigma_1\cdot\vec\sigma_2) \,.
\end{equation}

The complete in-medium baryon-baryon potential due to the baryon species \(B\) in the Fermi sea is then given by the sum of the contributions written in \eqs{med1d}--\eq*{med3e}, \eq*{med4d}--\eq*{med5b} and \eq*{med6}.

\subsection{In-medium lambda-nucleon interaction} \label{subsec:lni}

In view of its outstanding role in hypernuclear physics, we present here as an example the explicit expressions for the effective \(\Lambda N\) interaction in isospin-symmetric as well as isospin-asymmetric nuclear matter (\(\rho_p\neq\rho_n\)), as it results from two-pion-exchange, one-pion-exchange and contact \(\Lambda NN\) three-body forces.
Only the expressions for the \(\Lambda n\) potential {need to be given}. The \(\Lambda p\) potential can be easily written by interchanging the Fermi momenta \(k_f^p\) with \(k_f^n\) (or the densities \(\rho_p\) with \(\rho_n\)) in the expressions for \(\Lambda n\).
Note that this relation between the \(\Lambda n\) and \(\Lambda p\) potentials provides a non-trivial check of our calculation.

The following expressions result from summing up the contributions from the protons and neutrons in the Fermi sea.
Note that the topologies (1), (2a) and (2b) vanish here due to the non-existence of an isospin-symmetric \(\Lambda\Lambda\pi\) vertex.
Therefore, the two-pion exchange contribution to the effective \(\Lambda n\) potential stems solely from the topology (3) and it reads
\begin{align*} \label{eq:med3ln}
V^\mathrm{med,3,D}_{\Lambda n} ={}& -\frac{g_A^2}{12\pi^2f_0^4}\Bigg\{
(3 b_0 + b_D)\m^2 \frac12\big[2\tilde\Gamma_0(p)-(q^2+2\m^2)\tilde G_0(p,q)\big] \\ &
+ (2 b_2 + 3 b_4) \frac14\big[ \frac83 ({(k_f^n)}^3 + 2 {(k_f^p)}^3) - 4(q^2+2\m^2)\tilde\Gamma_0(p) - 2q^2 \tilde\Gamma_1(p) + (q^2+2\m^2)^2\tilde G_0(p,q) \big]  \\ &
+ \mathrm i (\vec q\times\vec p\,)\cdot\vec\sigma_2 \Big[ (3 b_0 + b_D)\m^2(\tilde G_0(p,q)+2\tilde G_1(p,q)) +
\frac12 (2 b_2 + 3 b_4) \big(2\tilde\Gamma_0(p)+2\tilde\Gamma_1(p) \\
&\qquad -(q^2+2\m^2)(\tilde G_0(p,q)+2\tilde G_1(p,q))\big) \Big] \Bigg\} \,, \numberthis
% V^\mathrm{med,3,E}_{\Lambda n} ={}& 0 \,, \numberthis
\end{align*}
where we have introduced the linear combinations \(\tilde\Gamma_i(p)=\Gamma_i(p,k_f^n)+2\Gamma_i(p,k_f^p)\) and \(\tilde G_i(p,q)=G_i(p,q,k_f^n)+2G_i(p,q,k_f^p)\).
Note that the exchange contribution vanishes identically (in the case of two-pion exchange), \(V^\mathrm{med,3,E}_{\Lambda n}=0\). 
The potential in \eq{eq:med3ln} depends on the axial-vector coupling constant $g_A$ and on several LECs ($b$'s) of the sub-leading chiral
meson-baryon Lagrangian \ct*{Krause1990,Frink2004,Oller2006,Mai2009}. 
The only spin-dependent term is the one proportional to \(\vec\sigma_2 = \frac12(\vec\sigma_1+\vec\sigma_2)-\frac12(\vec\sigma_1-\vec\sigma_2)\) and therefore one recognizes a symmetric and an antisymmetric spin-orbit potential of equal but opposite strength.
Note that the in-medium $NN$ potential due to two-pion exchange possesses a much richer spin structure, cf.\ Appendix A.

Interestingly, topology (4) gives rise to a one-pion exchange \(\Lambda n\) interaction,
\begin{equation}
V^{\mathrm{med},4,D}_{\Lambda n} = \frac{\ld'_1g_A(\rho_p-\rho_n)}{2f_0^2(q^2+\m^2)}\vec\sigma_1\cdot\vec q\ \vec\sigma_2\cdot\vec q \,, \qquad V^{\mathrm{med},4,E}_{\Lambda n} = 0 \,,
\end{equation}
which is induced by an isospin-asymmetry in the nuclear medium.
Again, there is no contribution from the exchange-type diagrams.
Furthermore, the topologies (5a) and (5b) lead to the combined in-medium \(\Lambda n\) potential:
\begin{align*}
V^{\mathrm{med},5,a+b}_{\Lambda n} = \frac{g_A}{4\pi^2f_0^2} \bigg\{ &
\frac{\ld'_2}3\big(2({(k_f^n)}^3+2{(k_f^p)}^3)-3\m^2\tilde\Gamma_0(p)\big) + \ld'_1 \tilde\Gamma_2(p)\vec\sigma_1\cdot\vec\sigma_2 \\ &
+ \ld'_1 \frac{\vec\sigma_1\cdot\vec p \ \vec\sigma_2\cdot\vec p + \vec\sigma_1\cdot\vec p^{\,\prime} \ \vec\sigma_2\cdot\vec p^{\,\prime} }2
\big(\tilde\Gamma_0(p)+2\tilde\Gamma_1(p)+\tilde\Gamma_3(p)\big)
\bigg\} \,, \numberthis
\end{align*}
where the expressions for the constants \(\ld'_1\) and \(\ld'_2\) can be found in Eq.~(44) of \ct{Petschauer2016}.

Finally, one obtains from the \(\Lambda NN\) contact interaction the following contribution to the in-medium potential
\begin{equation}
V^{\mathrm{med},6}_{\Lambda n} =
4\rho_n \lc'_1
+2\rho_p\big(\lc'_1+3\lc'_3+\lc'_2\,\vec\sigma_1\cdot\vec\sigma_2\big) \,,
\end{equation}
where the definition of the constants \(\lc'_i\) can be found in Eq.~(39) of \ct{Petschauer2016}.

Moreover, as a check we have rederived the effective $NN$ interaction in isospin-symmetric nuclear matter within our formalism which already includes antisymmetrization.
The corresponding results are summarized in \app{app:NN} and these agree with the antisymmetrized expressions of \ct{Holt:2009ty}.

\section{Three-baryon force through decuplet saturation}  \label{sec:dec}

In this section an estimate is performed for the LECs of the leading 3BFs  by applying decuplet saturation.
This concerns the meson-baryon LECs $b_{0,F,D}$, $b_{1,2,3,4}$, $d_{1,2,3}$ 
that appear in the two-meson exchange 3BF and the
LECs $D_i$ and $C_i$ for the one-meson exchange and contact 3BFs \ct*{Petschauer2016}. 
We use these ``saturated'' LECs in the evaluation of the 
effective \(\Lambda n\) interaction presented in the previous section. The aim is to provide a qualitative assessment of its density dependence induced by the different pieces of three-body interactions.
This estimate via decuplet saturation actually applies beyond the present consideration of chiral 3BFs in nuclear matter 
to any few- or many-baryon system where three-body forces are of relevance. 

The estimated LECs are obtained by calculating the diagrams in \fig{fig:hierdec} including decuplet baryons as intermediate states. The chiral Lagrangian for the octet-to-decuplet baryon transition involving a single pseudoscalar meson is employed,
and the pertinent non-relativistic contact vertex between three octet baryons and one decuplet baryon, \(B^*BBB\), is constructed.
Note that in the nucleonic sector only the two-pion exchange diagram with an intermediate \(\Delta(1232)\)-isobar in \fig{fig:hierdec} is allowed.
Other diagrams are forbidden by the Pauli principle, as will be shown in \app{app:DBBB}.
In fact, for three flavors the corresponding group theoretical considerations restrict the number of possible contact couplings \(BB\to B^*B\) to only two.

\subsection{Lagrangians including decuplet baryons} \label{subsec:DecL}

Here, we present the minimal set of terms in the chiral Lagrangian, that are necessary for evaluating the diagrams including decuplet baryons in \fig{fig:hierdec}.
The leading-order interaction Lagrangian between octet and decuplet baryons and octet pseudoscalar mesons that respects SU(3) symmetry reads in the non-relativistic limit (see, \eg \ct{Sasaki2006}):
\begin{equation} \label{eq:LMBD}
 \mathscr{L} = \frac C{f_0} \sum_{a,b,c,d,e=1}^3 \epsilon_{abc} \left( \bar T_{ade} \vec S^{\,\dagger} \cdot (\vec\nabla \phi_{db})B_{ec} + \bar B_{ce} \vec S \cdot (\vec\nabla\phi_{bd})T_{ade} \right)\,,
\end{equation}
where the decuplet baryons are combined to the totally symmetric three-index tensor \(T\), with components
\begin{align*} \label{eq:Tfields}
  T^{111}&=\Delta^{++}\,, \quad& T^{112}&=\tfrac{1}{\sqrt{3}}\Delta^{+}\,, \quad& T^{122}&=\tfrac{1}{\sqrt{3}}\Delta^{0}\,, \quad& T^{222}&=\Delta^{-}\,, \\
  T^{113}&=\tfrac{1}{\sqrt{3}}\Sigma^{*+}\,, & T^{123}&=\tfrac{1}{\sqrt{6}}\Sigma^{*0}\,, & T^{223}&=\tfrac{1}{\sqrt{3}}\Sigma^{*-}\,, \\
  T^{133}&=\tfrac{1}{\sqrt{3}}\Xi^{*0}\,, & T^{233}&=\tfrac{1}{\sqrt{3}}\Xi^{*-}\,, \\
  T^{333}&=\Omega^-\,. \numberthis
\end{align*}
The traceless $3\times 3$ matrices $B$ and $\phi$ in \eq{eq:LMBD} include the octet baryons and the pseudo-scalar mesons 
and their explicit form can be found, \eg in \ct{Petschauer2016}. 
The spin transition matrices \(\vec S\) connect two-component spinors of octet baryons with four-component 
spinors of decuplet baryons (cf.\ \ct{Weise1988}), and they fulfill the relation \( S_i {S_j}^\dagger = \frac13 ( 2\delta_{ij}-\mathrm i\epsilon_{ijk} \sigma_k )\).
Only one single LEC, $C$, is present at leading order and we use for it the (large-\(N_c\)) value \(C=\frac34g_A\approx 1\) \ct*{Kaiser1998}.
Rewriting the lowest-order decuplet Lagrangian in \eq{eq:LMBD} in the particle basis gives%
\begin{equation}
\mathscr{L} = \frac C{f_0} \sum_{i,j,k} N_{B_i^*\phi_jB_k}
\left[
\bar B_i^* \vec S^{\,\dagger} \cdot \left(\vec\nabla \phi_j\right) B_k +
\bar B_k \vec S \cdot \left(\vec\nabla \phi_j^\dagger\right) B_i^*
\right] \,,
\end{equation}
with SU(3) coefficients \(N_{B_i^*\phi_jB_k}\) and with the physical meson fields
\(\phi_i \in \big\{\pi^0\), \(\pi^+\), \(\pi^-,K^+\), \(K^-\), \(K^0\), \(\bar K^0\), \(\eta\big\}\), octet baryon fields
\(B_i \in \big\{n\), \(p\), \(\Lambda\), \(\Sigma^0\), \(\Sigma^+\), \(\Sigma^-\), \(\Xi^0\), \(\Xi^-\big\}\) and decuplet baryon fields
\(B^*_i \in \big\{\Delta^-\), \(\Delta^0\), \(\Delta^+\), \(\Delta^{++}\), \(\Sigma^{*0}\), \(\Sigma^{*+}\), \(\Sigma^{*-}\), \(\Xi^{*0}\), \(\Xi^{*-}\), \(\Omega^-\big\}\).

%%%

The other vertex including decuplet baryons that appears in \fig{fig:hierdec} is the leading-order \(B^*BBB\) contact vertex, involving three octet and one decuplet baryon.
The minimal non-relativistic contact Lagrangian that respects SU(3) symmetry takes the form (in matrix notation):
\begin{align*} \label{eq:LDBBBmin}
\mathscr{L} = &\quad\,
\lC_1\sum_{\substack{a,b,c,\\d,e,f=1}}^3 \epsilon_{abc}
\big[
\left(\bar T_{ade}\vec S^\dagger B_{db}\right)\cdot\left(\bar B_{fc}\vec\sigma B_{ef}\right)+
\left(\bar B_{bd}\vec S\, T_{ade}\right)\cdot\left(\bar B_{fe}\vec\sigma B_{cf}\right)
\big]\\
&+\lC_2
\sum_{\substack{a,b,c,\\d,e,f=1}}^3 \epsilon_{abc}
\big[
\left(\bar T_{ade}\vec S^\dagger B_{fb}\right)\cdot\left(\bar B_{dc}\vec\sigma B_{ef}\right)+
\left(\bar B_{bf}\vec S\, T_{ade}\right)\cdot\left(\bar B_{fe}\vec\sigma B_{cd}\right)
\big] \,, \numberthis
\end{align*}
with two low-energy constants \(\lC_1\) and \(\lC_2\).
The derivation of this minimal Lagrangian consistent with group theoretical considerations can be found in \app{app:DBBB}.
In the particle basis the Lagrangian in \eq{eq:LDBBBmin} reads:
\begin{equation}
\mathscr{L} = \sum_{\kappa=1}^2 \lC_\kappa \sum_{i,j,k,l} N^\kappa_{B_i^*B_jB_kB_l}
\Big[
\left(\bar B_i^*\vec S^\dagger B_j\right)\cdot\left(\bar B_k\vec\sigma B_l\right)+
\left(\bar B_j\vec S\, B_i^*\right)\cdot\left(\bar B_l\vec\sigma B_k\right)
\Big] \,,
\end{equation}
where \(i\) runs now over the decuplet baryons; \(j,k,l\) run over the octet baryons, and the \(N\)'s are again SU(3) coefficients.
According to naive dimensional counting \ct*{Friar1997}, the constants \(H_1\) and \(H_2\) should be of the order \(\mathcal O(1/f_0^2)\).
Note that this Lagrangian automatically incorporates the feature that \(NN\to \Delta N\) transitions in \(S\)-waves are forbidden by the Pauli exclusion principle.
Isospin conservation requires that the \(NN\) state has total isospin \(I=1\), but then the total spin is \(S=0\) which is impossible in the \(N\Delta\) system.

\subsection{Estimates of low-energy constants} \label{subsec:est}

Now we estimate the LECs of the leading three-baryon interaction by evaluating the diagrams with intermediate decuplet baryons shown 
in \fig{fig:hierdec}.
By comparison with the general three-baryon potentials one can directly read off the LECs at leading order. As a by-product one obtains a set of relations between these constants.

%%%

\begin{figure}[ht!]
\centering
\(
\vcenter{\hbox{\includegraphics[scale=.6]{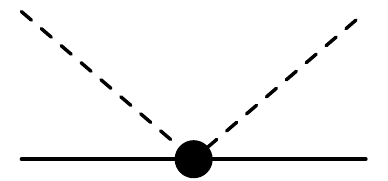}}}
\quad = \quad
\vcenter{\hbox{\includegraphics[scale=.6]{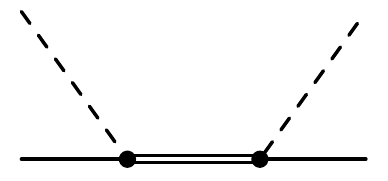}}}
\quad + \quad
\vcenter{\hbox{\includegraphics[scale=.6]{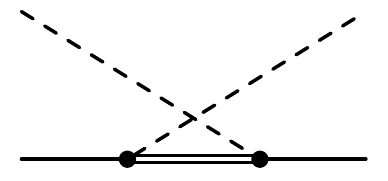}}}
\)
\caption{Saturation of the NLO baryon-meson vertex via decuplet resonances.}
\label{fig:3BF2MEDec}
\end{figure}
For the three-baryon interaction with two-meson exchange it suffices to consider the subprocess \(B_1\phi_1\to B_2\phi_2\) shown in \fig{fig:3BF2MEDec}.
The general diagram on the left-hand side stems from the Lagrangian Eq.~(31) in \ct{Petschauer2016} and it provides the following transition matrix element%
\begin{align*}
 V ={}&\phantom{+\ } \sum_{c^f=b_D,b_F,b_0}\frac{c^f}{4f_0^2} ( N^f_{\phi_1\substack{o\\i}\bar\phi_2} + N^f_{\bar\phi_2\substack{o\\i}\phi_1}) \\
 &+ \sum_{c^f=b_1,b_2,b_3,b_4}\frac{c^f}{f_0^2} ( N^f_{\phi_1\substack{o\\i}\bar\phi_2} + N^f_{\bar\phi_2\substack{o\\i}\phi_1})\vec q_1\cdot\vec q_2 \\
 &- \sum_{c^f=d_1,d_2,d_3}\frac{c^f}{f_0^2} ( N^f_{\phi_1\substack{o\\i}\bar\phi_2} - N^f_{\bar\phi_2\substack{o\\i}\phi_1})\,\mathrm i\, (\vec q_1\times\vec q_2)\cdot\vec\sigma \,,\numberthis
\end{align*}
where the SU(3) coefficients \(N^f\) are defined in Eq.~(33) of \ct{Petschauer2016}.
From the two diagrams on the right-hand side of \fig{fig:3BF2MEDec} (with intermediate decuplet baryons) one obtains
\begin{align*}
V = -\frac{C^2}{3\Delta f_0^2} \big[\,
&2(N_{B^* \phi_2 B_o}N_{B^* \phi_1 B_i} + N_{B^* \bar \phi_1 B_o}N_{B^* \bar\phi_2 B_i})\,\vec q_1\cdot\vec q_2 \\
&+(N_{B^* \phi_2 B_o}N_{B^* \phi_1 B_i} - N_{B^* \bar \phi_1 B_o}N_{B^* \bar\phi_2 B_i})\,\mathrm i\,(\vec q_1\times\vec q_2)\cdot\vec\sigma\,
\big]\,, \numberthis
\end{align*}
where we have introduced the average decuplet-octet baryon mass splitting \(\Delta= M_{10}-M_{8}\).
After summing over all intermediate decuplet baryons \(B^*\), a direct comparison of the transition matrix elements for all combinations of baryons and mesons leads
to the following relations for the LECs of the meson-baryon Lagrangian in Eq.~(31) of \ct{Petschauer2016}:
\begin{align*}
b_D  &= 0 \,,\
b_F  = 0 \,,\
b_0  = 0 \,,\\
b_1  &= \frac{7 C^2}{36 \Delta } \approx 0.59 \,,\
b_2  = \frac{C^2}{4 \Delta } \approx 0.76 \,,\
b_3  = -\frac{C^2}{3 \Delta } \approx -1.01 \,,\
b_4  = -\frac{C^2}{2 \Delta } \approx -1.51 \,,\\
d_1  &= \frac{C^2}{12 \Delta } \approx 0.25 \,,\
d_2  = \frac{C^2}{36 \Delta } \approx 0.08 \,,\
d_3  = -\frac{C^2}{6 \Delta } \approx -0.50 \,, \numberthis
\end{align*}
where all numerical values are in \(\mathrm{GeV}^{-1}\) and we have inserted \(\Delta\approx 300\ \mathrm{MeV}\) together with \(C=\frac34g_A\approx0.95\).
According to dimensional arguments \ct*{Friar1997,Epelbaum2002} the constants \(b_i,d_i\) are of order \(\mathcal O(1/\Lambda_\chi)\), 
with \(\Lambda_\chi\) the chiral symmetry breaking scale of the order of 1 GeV.
These constants are formally enhanced by a factor \(\Lambda_\chi/\Delta\) and thus promoted to \(\mathcal O(1/\Delta)\).
Obviously all LECs are proportional to \(C^2\), so that the two-meson exchange 3BF does not involve any unknown constant in decuplet saturation.
The result above is in line with the well-known \(\Delta\)(1232) contributions to the LECs \(c_1,c_3,c_4\) in the nucleonic sector \ct*{Bernard1997,Frink2004,Epelbaum2008a}:
\begin{equation}
c_1=\frac12(2b_0+b_D+b_F)=0\,,\qquad c_3= b_1+b_2+b_3+2b_4=-\frac{g_A^2}{2 \Delta } \,,\qquad c_4=4(d_1+d_2)=\frac{g_A^2}{4\Delta } \,.
\end{equation}
%%%

\begin{figure}[ht]
\begin{minipage}[b]{.35\textwidth}
\centering
\(
\vcenter{\hbox{\includegraphics[scale=.6]{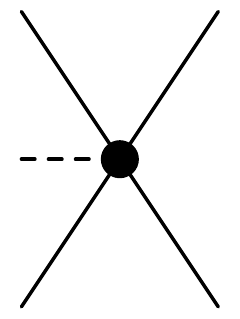}}} =
\vcenter{\hbox{\includegraphics[scale=.6]{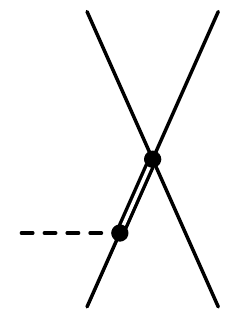}}} +
\vcenter{\hbox{\includegraphics[scale=.6]{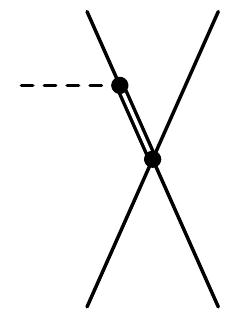}}}
\)
\vspace{1.2\baselineskip}
\caption{Saturation of the \(BB\to BB\phi\) vertex via decuplet resonances.}
\label{fig:3BF1MEDec}
\end{minipage}
\hfill\hfill
\begin{minipage}[b]{.24\textwidth}
\vspace{.3\baselineskip}
\centering
\begin{overpic}[scale=.6]{FvertBBMBB}
\put(0,103){$k$}\put(67,103){$l$}
\put(0,-15){$i$}\put(67,-15){$j$}
\put(-5,-37){$A$}\put(61,-37){$B$}
\put(-12,46){$\phi$}
\end{overpic}
\vspace{1.2\baselineskip}
\caption{Generic \(BB\to BB\phi\) diagram.}
\label{fig:3BF1MEGen}
\end{minipage}
\hfill\hfill
\begin{minipage}[b]{.32\textwidth}
\vspace{\baselineskip}
\centering
\(
\vcenter{\hbox{\begin{overpic}[scale=.6]{FvertBBDMBB}
\put(24,103){$k$}\put(65,103){$l$}
\put(24,-15){$i$}\put(65,-15){$j$}
\put(18,-37){$A$}\put(59,-37){$B$}
\put(-12,22){$\phi$}
\put(18,38){$B^*$}
\end{overpic}}}
\ + \quad
\vcenter{\hbox{\begin{overpic}[scale=.6]{FvertBBMDBB}
\put(24,103){$k$}\put(65,103){$l$}
\put(24,-15){$i$}\put(65,-15){$j$}
\put(18,-37){$A$}\put(59,-37){$B$}
\put(-12,68){$\phi$}
\put(14,48){$B^*$}
\end{overpic}}}
\)
\vspace{1.2\baselineskip}
\caption{Generic \(BB\to BB\phi\) decuplet diagrams.}
\label{fig:3BF1MEDecGen}
\end{minipage}
\end{figure}

\pagebreak[1]

%%%

Now we turn to the one-meson-exchange part of the three-baryon forces.
It is sufficient to study decuplet saturation for the subprocess \(B_1B_2\to B_3B_4\phi\) depicted in \fig{fig:3BF1MEDec}.
In the generic diagram of \fig{fig:3BF1MEGen} the baryon pairs \(i\)-\(k\) and \(j\)-\(l\) are in spin-spaces \(A\) and \(B\), respectively.
The corresponding transition amplitude reads
\begin{equation}
V^{AB}_{\substack{kl\\ij}} = \frac{\mathrm i}{f_0}\bigg(
\sum_{f=1}^{10} \ld_f N^{f}_{\substack{kl\\ij}\,\bar\phi}\ \vec\sigma_B\cdot \vec q  +
\sum_{f=11}^{14} \ld_f N^{f}_{\substack{kl\\ij}\,\bar\phi}\ \mathrm i\,(\vec\sigma_A\times\vec\sigma_B)\cdot\vec q
\bigg) \,,
\end{equation}
where \(\vec q\) is the momentum of the emitted meson and the \(N^f\) are the SU(3) coefficients defined in Eq.~(28) of \ct{Petschauer2016}.
Since the baryon \(B_1\) in the initial state belongs (per definition) to spin-space \(1\) and \(B_2\) to spin-space \(2\), the labels \(A\) and \(B\) 
are determined by \(i\) and \(j\), and, therefore, we can drop them in the notation. %can define \(V_{\substack{ij\\kl}} = V^{ij}_{\substack{ij\\kl}}\).
The complete transition matrix element of the process \(B_1B_2\to B_3B_4\phi\) is then given by two direct diagrams and two exchanged diagrams to which the (negative) spin-exchange operator \(P^{(\sigma)}=\frac12(\mathbbm{1}+\vec\sigma_1\cdot\vec\sigma_2)\) has to be applied in the final state:%
\begin{equation} \label{eq:BBMBBfull}
V = V_{\substack{34\\12}} + V_{\substack{43\\21}} - P^{(\sigma)} \bigg( V_{\substack{43\\12}} + V_{\substack{34\\21}} \bigg) \,.
\end{equation}
In the next step we consider the two diagrams on the right-hand side of \fig{fig:3BF1MEDec}.
From the generic diagrams in \fig{fig:3BF1MEDecGen} including a decuplet baryon \(B^*\) one finds the following transition matrix element (with yet unspecified spin spaces \(A\) and \(B\)):
\begin{align*} \label{eq:BBMBBDecGen}
V^{AB}_{\substack{kl\\ij}} ={}& \frac{\mathrm i\,C}{3\Delta f_0}\Big[
(\lC_1 N^1_{B^*B_{k}B_{j}B_{l}} + \lC_2 N^2_{B^*B_{k}B_{j}B_{l}}) N_{B^* \bar \phi B_i}
\big(2\vec\sigma_B\cdot\vec q-\mathrm i (\vec\sigma_A\times\vec\sigma_B)\cdot\vec q\,\big)\\
&\qquad\quad+(\lC_1 N^1_{B^*B_{i}B_{l}B_{j}} + \lC_2 N^2_{B^*B_{i}B_{l}B_{j}}) N_{B^* \phi B_k}
\big(2\vec\sigma_B\cdot\vec q+\mathrm i (\vec\sigma_A\times\vec\sigma_B)\cdot\vec q\,\big)\Big] \,. \numberthis
\end{align*}
It gets completed by antisymmetrization according to \eq{eq:BBMBBfull} and summing over all intermediate decuplet baryons \(B^*\).
Now, we can compare the complete transition matrix elements for all combinations of baryons and mesons.
This leads to the following results for the 14 LECs of the minimal non-relativistic chiral Lagrangian for the four-baryon-one-meson contact vertices (see Eq.~(27) in \ct{Petschauer2016}):
%\begin{multicols}{2}
\vspace{\baselineskip}
\twocolumngrid
\allowdisplaybreaks
\noindent
\begin{align*}
\input{files/BBMBBDecconstants}
\end{align*}
%\end{multicols}
\onecolumngrid
\vspace{\baselineskip}
According to dimensional arguments \ct*{Friar1997,Epelbaum2002} the constants \(D_i\) are of order \(\mathcal O(1/(\Lambda_\chi f_0^2))\).
In the nucleonic sector the corresponding single constant \(D=4(D_1-D_3+D_8-D_{10})\) is commonly denoted by \(D=c_D/(\Lambda_\chi f_0^2)\), 
where $c_D$ is of order $1$.
In the decuplet approximation for three flavors the constants \(D_i\) get
promoted to order \(\mathcal O(1/(\Delta f_0^2))\).

%%%

\begin{figure}[htbp!]
\mbox{}\hfill
\begin{minipage}[b]{.47\textwidth}
\centering
\(
\vcenter{\hbox{\includegraphics[scale=.6]{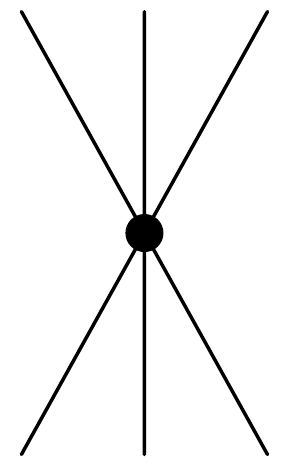}}}
\ =\
\vcenter{\hbox{\includegraphics[scale=.6]{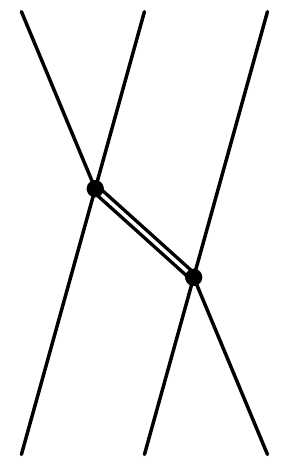}}}
\)
\caption{Saturation of the six-baryon contact interaction via decuplet resonances.}
\label{fig:3BFctDec}
\end{minipage}
\hfill\hfill
\begin{minipage}[b]{.4\textwidth}
\vspace{1.0\baselineskip}
\centering
\begin{overpic}[scale=.6]{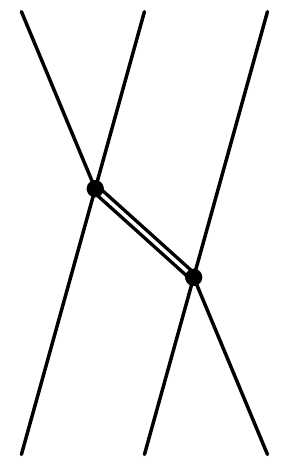}
\put(2,103){$l$}\put(27,103){$m$}\put(55,103){$n$}
\put(2,-10){$i$}\put(27,-10){$j$}\put(55,-10){$k$}
\put(0,-25){$A$}\put(25,-25){$B$}\put(53,-25){$C$}
\put(28,59){$B^*$}
\end{overpic}
\vspace{1.2\baselineskip}
\caption{Generic three-body decuplet contact diagram.}
\label{fig:3BFctDecGen}
\end{minipage}
\hfill\mbox{}
\end{figure}
\pagebreak[1]
Finally, in order to estimate the LECs of the six-baryon contact Lagrangian in Eq.~(14) of \ct{Petschauer2016}, we consider the process \(B_1B_2B_3\to B_4B_5B_6\) 
and its diagrammatic realization in terms of $B^*$ exchange, as shown in \fig{fig:3BFctDec}.
The three-body potential provided by the left diagram in \fig{fig:3BFctDec} is calculated by performing all 36 Wick contractions as described in Eqs.~(8) and (12) of \ct{Petschauer2016}.
The evaluation of the diagram on the right-hand side of \fig{fig:3BFctDec} follows using a similar procedure.
From the generic diagram in \fig{fig:3BFctDecGen} with an intermediate decuplet baryon, \(B^*\), in which the baryon pair \(i\)-\(l\) belongs to spin space \(A\), the pair \(j\)-\(m\) to spin space \(B\) and the pair \(k\)-\(n\) to spin space \(C\), one obtains the following expression for the transition potential
\begin{align*} \label{eq:decBBB}
V^{ABC}_{\substack{lmn\\ijk}}={}&-\frac{1}{3\Delta}
(\lC_1 N^1_{B^*B_{m}B_{i}B_{l}} + \lC_2 N^2_{B^*B_{m}B_{i}B_{l}})
(\lC_1 N^1_{B^*B_{j}B_{n}B_{k}} + \lC_2 N^2_{B^*B_{j}B_{n}B_{k}})\\
&\quad\times(2\vec\sigma_A\cdot\vec\sigma_C-\mathrm i (\vec\sigma_A\times\vec\sigma_C)\cdot\vec\sigma_B)\,. \numberthis
\end{align*}
Note that the relation for the transition spin operators, \( S_a {S_b}^\dagger = \frac13 ( 2\delta_{ab}-\mathrm i\epsilon_{abc} \sigma_c )\), implies a sum over the four decuplet spin states.
In order that \eq{eq:decBBB} becomes comparable to the three-baryon contact potential derived in \ct{Petschauer2016}, one still has to permute the three spin-1/2 fermions in the initial and in the final state (i.e.\ 36 Wick contractions have to be performed).
Since the baryons \(B_1,B_2,B_3\) are (per definition) in the spin-spaces \(1,2,3\), respectively, the assignments \(A,B,C\) are determined by \(i,j,k\), hence these superscripts can be dropped.
The six direct Wick contractions contributing to the process \(B_1B_2B_3\to B_4B_5B_6\) lead to the intermediate result
\begin{equation}
 V^D = V_{\substack{456\\123}} + V_{\substack{564\\231}} + V_{\substack{645\\312}} + V_{\substack{465\\132}} + V_{\substack{654\\321}} + V_{\substack{546\\213}} \,,
\end{equation}
and then the full potential comprising all 36 Wick contractions is obtained by applying to \(V^D\) further (sign-weighted) spin and particle exchanges (see Eq.~(12) of \ct{Petschauer2016}):
\begin{align*}
 V &= V^D
 + \Ps_{23}\Ps_{13} \Big(V^D\Big)_{\substack{4\to5\\5\to6\\6\to4}}
 + \Ps_{23}\Ps_{12} \Big(V^D\Big)_{\substack{4\to6\\5\to4\\6\to5}} \\
 &\quad\ - \Ps_{23} \Big(V^D\Big)_{\substack{4\to4\\5\to6\\6\to5}}
 - \Ps_{13} \Big(V^D\Big)_{\substack{4\to6\\5\to5\\6\to4}}
 - \Ps_{12} \Big(V^D\Big)_{\substack{4\to5\\5\to4\\6\to6}} \,. \numberthis
\end{align*}
After summing over all intermediate decuplet baryons \(B^*\), one compares the decuplet expressions with the full three-body contact potential for all possible combinations of six baryons.
The following results are found for the 18 LECs in the six-baryon contact Lagrangian (see Eq.~(14) of \ct{Petschauer2016}):
\vspace{.5\baselineskip}
\twocolumngrid
%\begin{multicols}{2}
\allowdisplaybreaks
\noindent
\begin{align*}
\input{files/BBBDecconstants}
\end{align*}
%\end{multicols}
\onecolumngrid
\vspace{.5\baselineskip}
Again, from dimensional scaling arguments the constants \(C_i\) should be of order \(\mathcal O(1/(\Lambda_\chi f_0^4))\).
In the nucleonic sector the corresponding constant \(E=2(C_4-C_9)\) is commonly denoted by \(E=c_E/(\Lambda_\chi f_0^4)\), where $c_E$ is of
order $1$. The decuplet saturation mechanism promotes the constants \(C_i\) to \(\mathcal O(1/(\Delta f_0^4))\).

%%%

In order to elucidate the pattern of decuplet saturation, we display in \tab{tab:3BFDec} the channels which are active in producing 3BFs for the $S=0$ and $-1$ sectors.
The decuplet resonances which occur as intermediate states are indicated explicitly for the three classes of three-body forces.
The transitions for strangeness \(-1\) are mostly saturated by the \(\Sigma^*(1385)\) resonance alone.
However, for some transitions involving pions also the \(\Delta(1232)\) isobar contributes.
Resonances with higher strangeness can not be reached.
Note that in contrast to the \(NNN\) interaction, for $S=-1$ the one-meson exchange and the contact 3BF also
receive contributions from the excitation of decuplet baryons.

\begin{table}
\centering
\mbox{}\hfill
\begin{tabular}[t]{>{$}c<{$}>{$}c<{$}>{$}c<{$}}
\toprule
\text{transition} & \text{type} & B^* \\\midrule
NNN\to NNN & \pi\pi & \Delta \\
\midrule
\Lambda NN\to\Lambda NN & \pi\pi & \Sigma^* \\
\Lambda NN\to\Lambda NN & \pi K & \Sigma^* \\
\Lambda NN\to\Lambda NN & KK & \Sigma^* \\
\Lambda NN\to\Lambda NN & \pi & \Sigma^* \\
\Lambda NN\to\Lambda NN & K & \Sigma^* \\
\Lambda NN\to\Lambda NN & \text{ct} & \Sigma^* \\
\midrule
\Lambda NN\leftrightarrow\Sigma NN & \pi\pi & \Delta,\Sigma^* \\
\Lambda NN\leftrightarrow\Sigma NN & \pi K & \Delta,\Sigma^* \\
\Lambda NN\leftrightarrow\Sigma NN & \pi\eta & \Sigma^* \\
\Lambda NN\leftrightarrow\Sigma NN & KK & \Sigma^* \\
\Lambda NN\leftrightarrow\Sigma NN & K\eta & \Sigma^* \\
\Lambda NN\leftrightarrow\Sigma NN & \pi & \Delta,\Sigma^* \\
\Lambda NN\leftrightarrow\Sigma NN & K & \Sigma^* \\
\Lambda NN\leftrightarrow\Sigma NN & \eta & \Sigma^* \\
\Lambda NN\leftrightarrow\Sigma NN & \text{ct} & \Sigma^* \\
\bottomrule
\end{tabular}
\hfill
\begin{tabular}[t]{>{$}c<{$}>{$}c<{$}>{$}c<{$}}
\toprule
\text{transition} & \text{type} & B^* \\\midrule
\Sigma NN\to\Sigma NN & \pi\pi & \Delta,\Sigma^* \\
\Sigma NN\to\Sigma NN & \pi K & \Delta,\Sigma^* \\
\Sigma NN\to\Sigma NN & \pi\eta & \Sigma^* \\
\Sigma NN\to\Sigma NN & KK & \Sigma^* \\
\Sigma NN\to\Sigma NN & K\eta & \Sigma^* \\
\Sigma NN\to\Sigma NN & \eta\eta & \Sigma^* \\
\Sigma NN\to\Sigma NN & \pi & \Delta,\Sigma^* \\
\Sigma NN\to\Sigma NN & K & \Sigma^* \\
\Sigma NN\to\Sigma NN & \eta & \Sigma^* \\
\Sigma NN\to\Sigma NN & \text{ct} & \Sigma^* \\
\bottomrule
\end{tabular}
\hfill\mbox{}
\caption{Enhanced three-body interactions through decuplet saturation for strangeness \(0\) and \(-1\) systems, with classes of diagrams as specified: two-meson exchange, one-meson exchange and contact interaction (ct).}
\label{tab:3BFDec}
\end{table}

\subsection{Lambda-nucleon-nucleon in decuplet approximation}

Using the LECs derived from decuplet saturation, this fixes the constants of the \(\Lambda NN\) (contact, one-pion and two-pion exchange) three-body interaction introduced in \ct{Petschauer2016}.
These particular linear combinations of coefficients read
\begin{alignat*}{2}
&\lc^\prime_1  = \lc^\prime_3 = {} \frac{\lC'^2}{72 \Delta } \,, \qquad\quad
&&\lc^\prime_2  ={} 0 \,, \\
&\ld'_1  ={} 0 \,, \qquad\quad
&&\ld'_2  ={} \frac{2\,C \lC'}{9 \Delta } \,, \\
&3b_0+b_D ={} 0 \,, \qquad\quad
&&2b_2+3b_4 ={} -\frac{C^2}{\Delta } \,, \numberthis
\end{alignat*}
and they depend only on the combination \(\lC' = \lC_1+3 \lC_2\) of the \(B^*BBB\) contact couplings.
Notably, the constants \(\lc^\prime_i\) of the \(\Lambda NN\) contact interaction are positive independently of the values \(\lC_1\) and \(\lC_2\).

With the above values estimated via decuplet saturation, the three components of the density-dependent \(\Lambda n\) potential in a nuclear medium with densities \(\rho_p\) and \(\rho_n\) take the following simple forms
\begin{align*} \label{eq:medDec2pe}
V^{\mathrm{med,\pi\pi}}_{\Lambda n} = \frac{C^2g_A^2}{12\pi^2f_0^4\Delta}\Bigg\{
&\frac14\big[ \frac83 ({k_f^n}^3 + 2 {k_f^p}^3) - 4(q^2+2m^2)\tilde\Gamma_0(p) - 2q^2 \tilde\Gamma_1(p) + (q^2+2m^2)^2\tilde G_0(p,q) \big]  \\
&+ \frac{\mathrm i}2 (\vec q\times\vec p\,)\cdot\vec\sigma_2
\big(2\tilde\Gamma_0(p)+2\tilde\Gamma_1(p)
-(q^2+2m^2)(\tilde G_0(p,q)+2\tilde G_1(p,q))\big) \Bigg\} \,, \numberthis
\end{align*}
\begin{align*} \label{eq:medDec1pe}
V^{\mathrm{med,\pi}}_{\Lambda n} = \frac{g_A C \lC'}{54\pi^2f_0^2\Delta}
\big(2({k_f^n}^3+2{k_f^p}^3)-3m^2\tilde\Gamma_0(p)\big) \,, \numberthis
\end{align*}
\begin{align} \label{eq:medDecCont}
V^{\mathrm{med,ct}}_{\Lambda n} &= \frac{\lC'^2}{18\Delta}(\rho_n +2\rho_p) \,,
\end{align}
where the different topologies related to two-pion exchange ((1), (2a), (2b), (3)) and one-pion exchange ((4), (5a), (5b)) have already been combined in $V{^\mathrm{med,\pi\pi}}$ and $V^{\mathrm{med,\pi}}$, respectively.
The density and momentum dependent functions \(\tilde\Gamma_i(p)\) and \(\tilde G_i(p,q)\) have been defined following \eq{eq:med3ln}.
Note that since \(D'_1\) vanishes in decuplet saturation, there remain only central and spin-orbit components for the \(\Lambda N\) in-medium potential.

\subsection{In-medium \texorpdfstring{{\boldmath \(\Sigma N\)}}{Sigma-N} interactions in decuplet approximation}

Other interesting parts of the in-medium potentials derived in \sect{sec:med} for the strangeness \(S=-1\) sector are those involving \(\Sigma N\) states.
Here we write down the explicit formulas for the corresponding (transition) potentials in isospin-symmetric nuclear matter, with density \(\rho=2\rho_n=2\rho_p=2k_f^3/(3\pi^2)\), employing the decuplet approximation.
In such a medium isospin symmetry still holds and it is sufficient to consider the potentials for the three independent transitions \(\Lambda N\to\Sigma N\) with isospin \(1/2\), \(\Sigma N\to\Sigma N\) with isospin \(1/2\), and \(\Sigma N\to\Sigma N\) with isospin \(3/2\).
The transformation from the particle basis to the isospin basis is performed as in Eq.~(19) of \ct{Petschauer2016}.
For more complicated cases such as hyperon-nucleon interactions in isospin-asymmetric nuclear matter or even in hypernuclear matter, one can use straightforwardly the general potential formulas given in \sect{sec:med} together with an automated calculation of the SU(3) coefficients, following the definitions in this work and in \ct{Petschauer2016}.

We restrict ourselves again to two-pion exchange, one-pion exchange and contact contributions to the in-medium potentials. Consequently all exchange-type contributions vanish, \(V^\mathrm{med,1,E}=V^\mathrm{med,2a,E}=V^\mathrm{med,2b,E}=V^\mathrm{med,3,E}=V^\mathrm{med,4,E}=0\), since these involve strangeness transfer from one baryon to the other.

After summing the contributions from the proton and neutron Fermi seas, the non-vanishing in-medium potentials for the three transitions \(\Lambda N\to\Sigma N\), \(\Sigma N\to\Sigma N\ (I=1/2)\), and \(\Sigma N\to\Sigma N\ (I=3/2)\), take the following form:
\begin{align*}
V^\mathrm{med,1,D}_{\Lambda N\to\Sigma N} &=
\frac{D}{2F}V^\mathrm{med,1,D}_{\Sigma N,1/2} =
-\frac{D}{F}V^\mathrm{med,1,D}_{\Sigma N,3/2} =
\frac{8\rho C^2 D g_A}{9 \Delta f_0^4} q^2 \frac{\vec\sigma_1\cdot\vec q\ \vec\sigma_2\cdot\vec q}{(q^2+\m^2)^2}  \,, \numberthis
\end{align*}

\begin{align*}
&V^\mathrm{med,2a+b,D}_{\Lambda N\to\Sigma N} =
\frac{D}{2F}V^\mathrm{med,2a+b,D}_{\Sigma N,1/2} =
-\frac{D}{F}V^\mathrm{med,2a+b,D}_{\Sigma N,3/2}  \\ &\quad =
-\frac{C^2 D g_A}{9 \Delta\pi^2f_0^4} \frac{\vec\sigma_1\cdot\vec q\ \vec\sigma_2\cdot\vec q}{q^2+\m^2} \Bigg\{
2 \Big[ \frac23k_f^3-\m^2\Gamma_0(p,k_f) + \Gamma_2(p,k_f) \Big] 
 +  \frac{q^2}2 \Big[\Gamma_0(p,k_f)+2\Gamma_1(p,k_f)+\Gamma_3(p,k_f)\Big]
\Bigg\} \,, \numberthis
\end{align*}

\begin{align*}
V^\mathrm{med,3,D}_{\Lambda N\to\Sigma N} ={}& -\frac{C^2 g_A^2}{12 \Delta\pi^2f_0^4} \Bigg\{ G_2(p,q,k_f) (q^2\vec\sigma_1\cdot\vec\sigma_2-\vec\sigma_1\cdot\vec q\ \vec\sigma_2\cdot\vec q\,) \\ &
+ \Big[G_0(p,q,k_f)+4G_1(p,q,k_f)+4G_3(p,q,k_f)\Big] (\vec q\times\vec p\,)\cdot\vec\sigma_1\, (\vec q\times\vec p\,)\cdot\vec\sigma_2 \\ &
- \frac{\mathrm i}2 (\vec q\times\vec p\,)\cdot\vec\sigma_1 \Big[
\Big(2\Gamma_0(p,k_f)+2\Gamma_1(p,k_f)-(q^2+2\m^2)\big(G_0(p,q,k_f)+2G_1(p,q,k_f)\big)\Big) \Big]
\Bigg\} \,, \\
V^\mathrm{med,3,D}_{\Sigma N,1/2} ={}& \frac{C^2 g_A^2}{36 \Delta\pi^2f_0^4} \Bigg\{ G_2(p,q,k_f) (q^2\vec\sigma_1\cdot\vec\sigma_2-\vec\sigma_1\cdot\vec q\ \vec\sigma_2\cdot\vec q\,) \\ &
+ \frac12\Big[ \frac83 k_f^3 - 4(q^2+2\m^2)\Gamma_0(p,k_f) - 2q^2 \Gamma_1(p,k_f) + (q^2+2\m^2)^2G_0(p,q,k_f) \Big] \\ &
+ \big[G_0(p,q,k_f)+4G_1(p,q,k_f)+4G_3(p,q,k_f)\big] (\vec q\times\vec p\,)\cdot\vec\sigma_1\, (\vec q\times\vec p\,)\cdot\vec\sigma_2 \\ &
- \frac{\mathrm i}2 (\vec q\times\vec p\,)\cdot(\vec\sigma_1-2\vec\sigma_2) \Big[
2\Gamma_0(p,k_f)+2\Gamma_1(p,k_f)-(q^2+2\m^2)\big(G_0(p,q,k_f)+2G_1(p,q,k_f)\big) \Big]
\Bigg\} \,, \\
V^\mathrm{med,3,D}_{\Sigma N,3/2} ={}& \frac{C^2 g_A^2}{72 \Delta \pi^2f_0^4} \Bigg\{ -G_2(p,q,k_f) (q^2\vec\sigma_1\cdot\vec\sigma_2-\vec\sigma_1\cdot\vec q\ \vec\sigma_2\cdot\vec q\,) \\ &
+ \frac83 k_f^3 - 4(q^2+2\m^2)\Gamma_0(p,k_f) - 2q^2 \Gamma_1(p,k_f) + (q^2+2\m^2)^2G_0(p,q,k_f) \\ &
- \big[G_0(p,q,k_f)+4G_1(p,q,k_f)+4G_3(p,q,k_f)\big] (\vec q\times\vec p\,)\cdot\vec\sigma_1\, (\vec q\times\vec p\,)\cdot\vec\sigma_2 \\ &
+ \frac{\mathrm i}2 (\vec q\times\vec p\,)\cdot(\vec\sigma_1+4\vec\sigma_2) \Big[ 2\Gamma_0(p,k_f)+2\Gamma_1(p,k_f)-(q^2+2\m^2)\big(G_0(p,q,k_f)+2G_1(p,q,k_f)\big) \Big]
\Bigg\} \,, \numberthis
\end{align*}

\begin{align*}
V^\mathrm{med,4,D}_{\Lambda N\to\Sigma N} &= \frac{2 \rho \lC_2 C g_A}{3 f_0^2 \Delta } \frac{\vec\sigma_1\cdot\vec q\ \vec\sigma_2\cdot\vec q}{q^2+\m^2} \,, \\
V^\mathrm{med,4,D}_{\Sigma N,1/2} &= \frac{4 \rho C g_A (2 \lC_1+\lC_2)}{9 f_0^2 \Delta } \frac{\vec\sigma_1\cdot\vec q\ \vec\sigma_2\cdot\vec q}{q^2+\m^2} \,, \\
V^\mathrm{med,4,D}_{\Sigma N,3/2} &= -\frac{2 \rho C g_A (2 \lC_1+\lC_2)}{9 f_0^2 \Delta } \frac{\vec\sigma_1\cdot\vec q\ \vec\sigma_2\cdot\vec q}{q^2+\m^2} \,, \numberthis
\end{align*}

\begin{align*}
V^{\mathrm{med},5a+b}_{\Lambda N\to\Sigma N} ={}& \frac{C g_A}{36\pi^2f_0^2\Delta} \bigg\{ 
\big(3 (\lC_1-\lC_2)-4(\lC_1+3 \lC_2)\vec\sigma_1\cdot\vec\sigma_2\big)\Big(\frac23k_f^3-\m^2\Gamma_0(p,k_f)\Big) 
+ 4\lC_1 \Gamma_2(p,k_f)\vec\sigma_1\cdot\vec\sigma_2 \\
&+ \Big( \frac{ 1}{4 }(19 \lC_1+9 \lC_2) \vec\sigma_1\cdot\vec p^{\,\prime} \ \vec\sigma_2\cdot\vec p^{\,\prime} -\frac{3 }{4  }(\lC_1+3 \lC_2) \vec\sigma_1\cdot\vec p \ \vec\sigma_2\cdot\vec p\, \Big) \big(\Gamma_0(p,k_f)+2\Gamma_1(p,k_f)+\Gamma_3(p,k_f)\big)
\bigg\} \,,  \\
V^{\mathrm{med},5a+b}_{\Sigma N,1/2} ={}& \frac{C g_A}{36\pi^2f_0^2\Delta} \bigg\{ 
\left(8 \lC_2 -2(3 \lC_1+\lC_2)\vec\sigma_1\cdot\vec\sigma_2\right)\Big(\frac23{k_f}^3-\m^2\Gamma_0(p,k_f)\Big) 
- 2(5 \lC_1+3 \lC_2) \Gamma_2(p,k_f)\vec\sigma_1\cdot\vec\sigma_2 \\
&  -( 5 \lC_1+3 \lC_2) \Big(  \vec\sigma_1\cdot\vec p^{\,\prime} \ \vec\sigma_2\cdot\vec p^{\,\prime} + \vec\sigma_1\cdot\vec p \ \vec\sigma_2\cdot\vec p\, \Big) \big(\Gamma_0(p,k_f)+2\Gamma_1(p,k_f)+\Gamma_3(p,k_f)\big)
\bigg\} \,,   \\
V^{\mathrm{med},5a+b}_{\Sigma N,3/2} ={}& \frac{C g_A}{36\pi^2f_0^2\Delta} \bigg\{ 
( 3 \lC_1+\lC_2)\left(2+\vec\sigma_1\cdot\vec\sigma_2\right)\Big(\frac23{k_f}^3-\m^2\Gamma_0(p,k_f)\Big)
+ (5 \lC_1+3 \lC_2) \Gamma_2(p,k_f)\vec\sigma_1\cdot\vec\sigma_2 \\
&+ \frac{ 1}{2 }(5 \lC_1+3 \lC_2) \Big(  \vec\sigma_1\cdot\vec p^{\,\prime} \ \vec\sigma_2\cdot\vec p^{\,\prime} + \vec\sigma_1\cdot\vec p \ \vec\sigma_2\cdot\vec p\, \Big) \big(\Gamma_0(p,k_f)+2\Gamma_1(p,k_f)+\Gamma_3(p,k_f)\big)
\bigg\} \,, \numberthis 
\end{align*}

\begin{align*}
V^{\mathrm{med},6}_{\Lambda N\to\Sigma N} &= \frac{\rho}{6\Delta} \left(\frac{1}{2 }(\lC_1^2+2 \lC_1 \lC_2-3 \lC_2^2)-\frac{1}{3 }(\lC_1^2+4 \lC_1 \lC_2+3 \lC_2^2)\vec\sigma_1\cdot\vec\sigma_2\right) \,, \\
V^{\mathrm{med},6}_{\Sigma N,1/2} &= \frac{\rho}{6\Delta}\left(-\frac{1}{2 }(\lC_1^2-2 \lC_1 \lC_2-7 \lC_2^2)+\frac{1}{3 }(\lC_1+\lC_2)^2\vec\sigma_1\cdot\vec\sigma_2\right) \,, \\
V^{\mathrm{med},6}_{\Sigma N,3/2} &= \frac{\rho}{6\Delta}\left(\frac{1}{2 }(5 \lC_1^2+2 \lC_1 \lC_2+\lC_2^2)-\frac{1}{6 }(\lC_1+\lC_2)^2\vec\sigma_1\cdot\vec\sigma_2\right) \,. \numberthis
\end{align*}
Note that the topologies (2a) and (2b), and (5a) and (5b), have already been combined.
One observes that these in-medium potentials exhibit a much richer spin-structure than the one for \(\Lambda N\to\Lambda N\).
Furthermore, the two constants \(\lC_1\) and \(\lC_2\) occur now in various combinations.

\section{Numerical results and discussion} \label{sec:res}

Selected numerical results are now presented for the in-medium $\Lambda N$ 
interaction based on the three contributions derived in the previous section. 
We restrict ourselves to the \(\Lambda n\) potential in isospin-symmetric nuclear matter and in pure neutron matter.
Obviously, the relation \(V^{\mathrm{med}}_{\Lambda p} = V^{\mathrm{med}}_{\Lambda n}\) holds generally in isospin-symmetric nuclear matter.
However, in decuplet approximation one deduces from \eqs{eq:medDec2pe}, \eq*{eq:medDec1pe} and \eq*{eq:medDecCont} for pure neutron matter the remarkable relation
\(V^{\mathrm{med}}_{\Lambda p} = 2V^{\mathrm{med}}_{\Lambda n}\).
For the presentation we follow closely \ct{Holt:2009ty} for the $NN$ case and show partial-wave projected momentum-space potentials (in units of \(\mathrm{fm}\), including a nucleon mass factor \(M_N\)).

In \ct{Holt:2009ty} the low-momentum potential \(V_{\rm low\, k}\) has been used for the free $NN$ interaction.
It is obtained from the bare chiral $NN$ potential at order N$^3$LO through evolution down to a low-momentum scale via renormalization
group techniques \ct*{Bogner2001,Bogner2003}.
At the chosen scale of $\Lambda_{\rm low \,k} = 2.1$~fm$^{-1}$ \ct*{Holt:2009ty}
basically all available high-precision $NN$ potentials converge to a nearly unique
low-momentum potential.
For the $\Lambda N$ case we do not have such an interaction at our disposal.
Though there are pertinent results in the literature \ct*{Schaefer:2005,Dapo:2008,Kohno2010}, 
it has to be said that there is no unique low-momentum potential in the $\Lambda N$ case because the 
relevant $\Lambda N$ phase shifts are not reliably known at present.
Different $YN$ potentials, fitted to the available scattering data, predict different phase shifts and thus yield different low-momentum potentials.
For the present exploratory study a bare $YN$ potential is used
and we take the NLO chiral interaction from \ct*{Haidenbauer2013a} with the lowest cutoff $\Lambda = 450$~MeV, close to $\Lambda_{\rm low \,k}$.
We expect that this bare interaction should be not too
far from a hypothetical ``universal potential'', at least for partial waves 
such as $^1S_0$, $^3P_0$, etc., \ie those without angular momentum mixing. The actual evaluation of $V_{\rm low\, k}$ 
for our LO $YN$ potential reported in \ct{Kohno2010} supports this expectation. 

\begin{figure}
\centering
\includegraphics[scale=0.55]{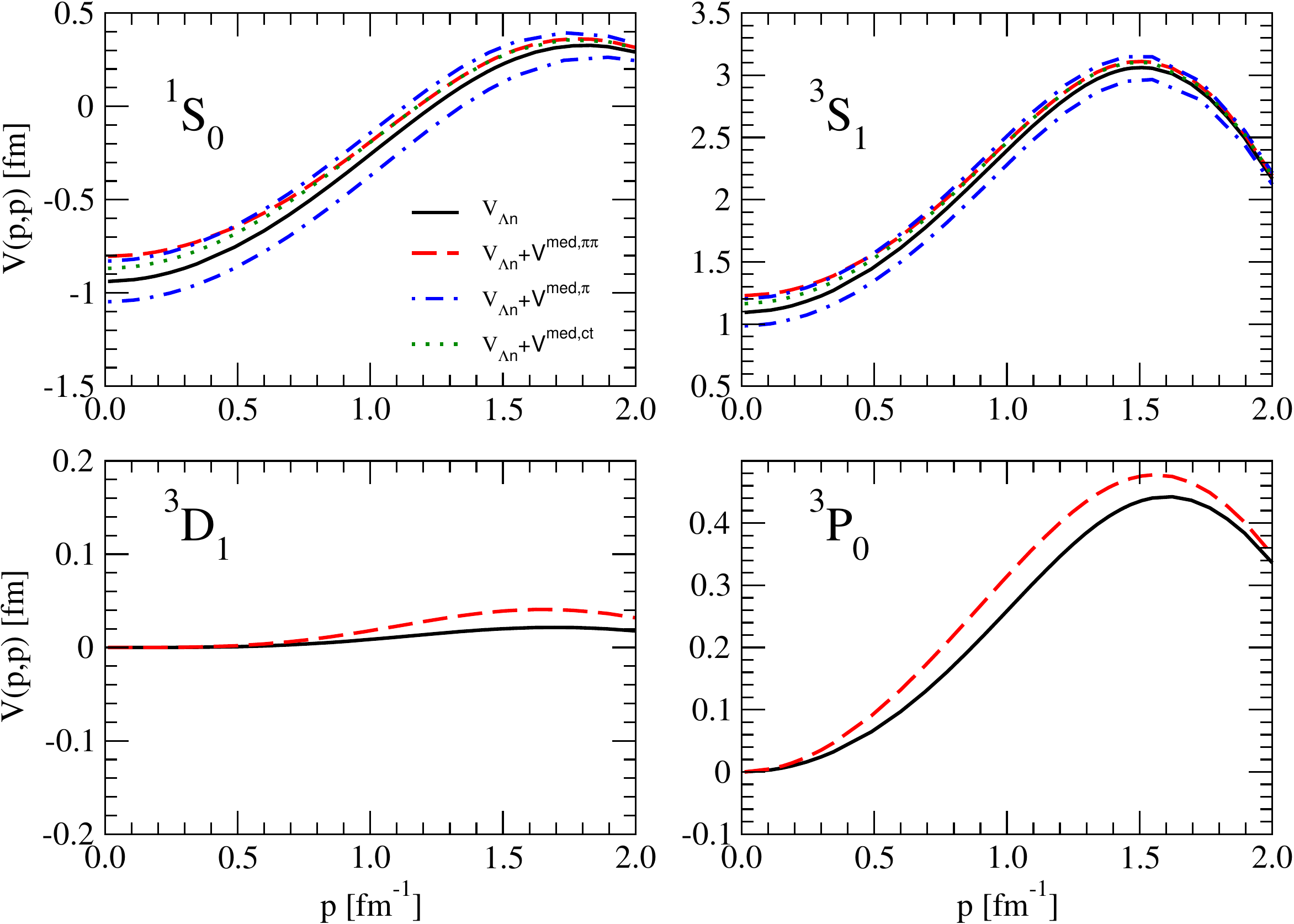}
\vspace{-.5\baselineskip}
\caption{Modifications of the (on-shell) $\Lambda n$  potential (solid/black line) 
due to the density-dependent contributions resulting from
the two-pion exchange (dashed/red line), one-pion exchange (dash-dotted/blue lines)
and contact (dotted/green line) three-body forces. 
The $YN$ potential at NLO in chiral EFT with cutoff $\Lambda = 450$ MeV 
\cite{Haidenbauer2013a} is used as basis. 
The two curves for one-pion exchange result from different signs of the
LEC $H'$, see text.
The calculations are for 
symmetric nuclear matter with $\rho = \rho_0 = 0.166$ fm$^{-3}$.
\label{fig:SNMs}
}
\vspace{.5\baselineskip}D
\includegraphics[scale=0.55]{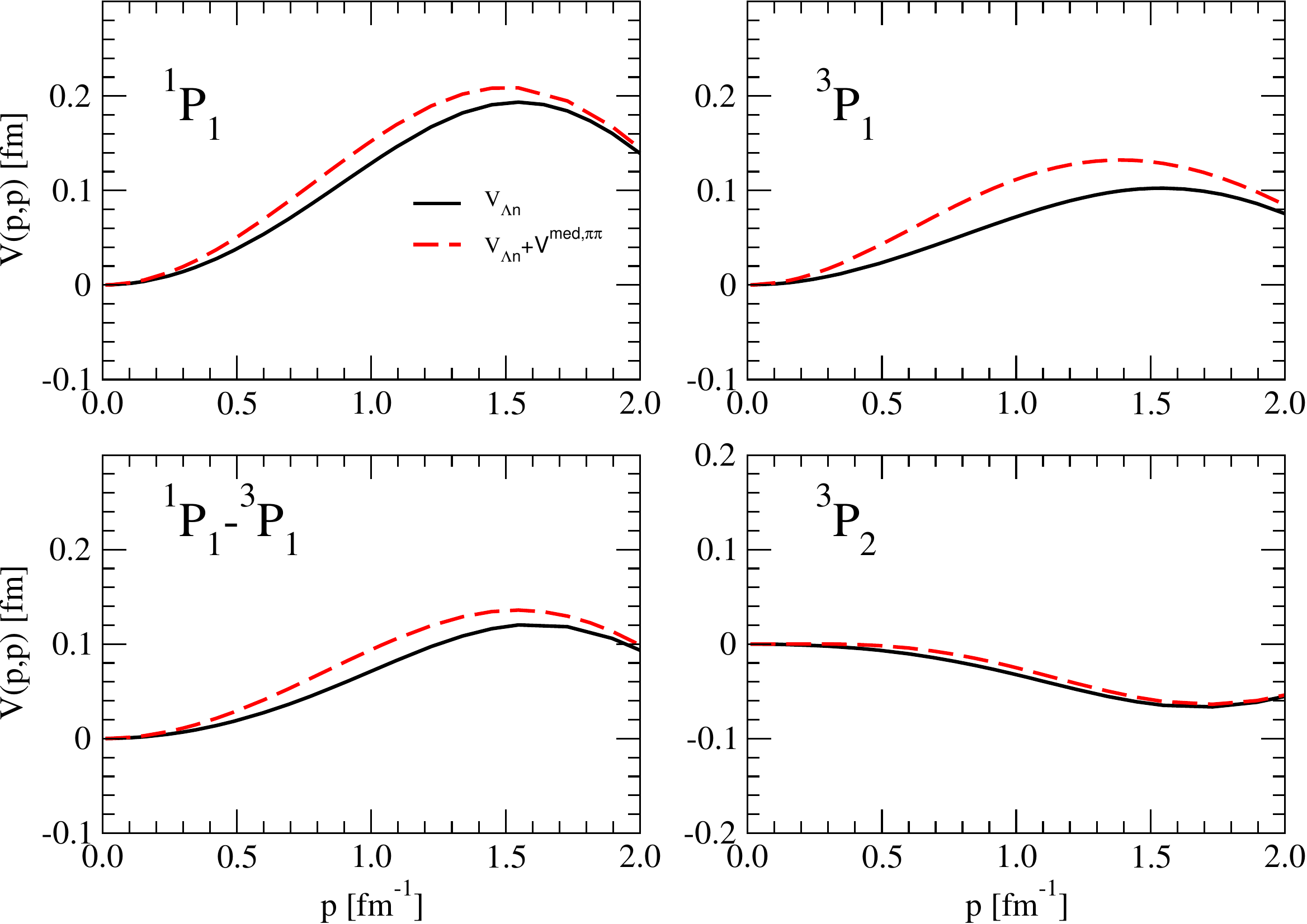}
\vspace{-.5\baselineskip}
\caption{
Modifications of the (on-shell) $\Lambda n$  potential (solid/black line)
due to the density-dependent contributions resulting from
three-body forces for selected \(P\)-waves. 
Same description as in Fig.~\ref{fig:SNMs}. 
\label{fig:SNMp}
}
\end{figure}
In Figs.~\ref{fig:SNMs} and \ref{fig:SNMp} the contributions of the density-dependent in-medium $\Lambda n$ 
interaction to the free space \(\Lambda n\) potential in low partial waves are displayed for isospin-symmetric nuclear 
matter at saturation density, $\rho_0=0.17$ fm$^{-3}$. The solid 
line represents the 
bare $\Lambda n$ potential while the dashed, dash-dotted, and dotted lines show the modifications 
due to the two-pion exchange ($\pi\pi$), one-pion exchange ($\pi$) and the contact ($\mathrm{ct}$) three-body force, 
respectively. 

Decuplet saturation fixes the parameters of the $\pi\pi$ contribution uniquely so that
the corresponding result can be considered as a prediction. The two other
contributions to the density-dependent effective $\Lambda n$ interaction depend on the unknown LEC $H'$. 
In the absence of more detailed information we assume that $H'\approx \pm 1/f_0^2$, in line with general dimensional scaling arguments \ct*{Friar1997,Epelbaum2002}. 
This is meant to be just a rough estimate.
One knows from the nucleonic case that the values of the LECs involved in the three-nucleon 
force ($c_D$ and $c_E$) depend strongly on the chosen scale and/or regularization scheme 
\ct*{Epelbaum2002,Nogga:2004}. In that case the LECs can be fixed by considering few-body observables 
such as $3N$ or $4N$ binding energies, Gamow-Teller matrix elements, etc. 
Whether a similar strategy can be followed for the hyperon sector by considering say
the $^{\,3}_\Lambda {\rm H}$ and $^{\,4}_\Lambda {\rm H}$ ($^{\,4}_\Lambda {\rm He}$) binding energies 
remains to be examined \ct*{PetschauerPrep}.

In any case, we can already draw some interesting qualitative conclusions.
First, two-pion exchange 3BFs lead to repulsive contributions in all partial waves. Moreover, the 
contact term also gives rise to overall repulsive contributions, independent of the uncertainty 
associated with its actual value. Here the decuplet saturation fixes the sign of the relevant
constants ($C_i'$) uniquely, as already mentioned in the preceding section. Only the
one-pion exchange contribution is sensitive to the sign of $H'$, and correspondingly it generates a repulsive
or attractive density-dependent $\Lambda n$ interaction, see the two dash-dotted curves in
\fig{fig:SNMs}. The particular choice $H'=|1/f^2_0|$ leads to results of
comparable magnitude for all three contributions. For a somewhat larger
value of the LEC $H'$ the contact interaction would dominate the density-dependent $\Lambda n$ interaction.
$H'$ enters quadratically in the corresponding potential $V^{\rm med,ct}_{\Lambda N}$,
cf.\ \eq{eq:medDecCont}, so that an increase of $H'$ by a factor of two (which is likewise in 
line with dimensional arguments) would enhance the corresponding contribution by a factor 4. 
 
Since $V^{\rm med,ct}_{\Lambda N}$ and $V^{\rm med,\pi}_{\Lambda N}$ do 
not depend on the momentum transfer $q$ they contribute only to $\Lambda N$ $S$-waves. Thus the 
uncertainty with regard to $H'$ does not affect the density-dependent interaction in the
$P$ waves (and other higher partial waves). With regard to our $P$-wave results, shown in
\fig{fig:SNMp}, it is of particular interest that $V^{\rm med,\pi\pi}_{\Lambda N}$ 
provides additional and repulsive contributions to the
antisymmetric spin-orbit force, see the $^1P_1$-$^3P_1$ transition amplitude.
As argued in \ct{Fujiwara2006,Haidenbauer2015a,Petschauer2015} based on $G$-matrix 
calculations of hyperons in nuclear matter, a sizable antisymmetric spin-orbit force that
can counterbalance the spin-orbit force generated by the basic interaction is one of
the possibilities to achieve a weak $\Lambda$-nucleus spin-orbit potential as 
indicated by experimental results for hypernuclear spectra \ct*{Hashimoto2006}.

Before comparing the density-dependent effects for $\Lambda N$ with those derived
for $NN$ in \ct{Holt:2009ty}, it should be noted that several 
topologies are absent in the former because the $\Lambda\Lambda\pi$ vertex does not exist. This
concerns specifically the one-pion exchange term with a Pauli blocked in-medium pion 
self-energy and vertex corrections to the one-pion exchange (topologies (1) and (2)) 
which provide the dominant density-dependent effects in the $NN$ case for on-shell
momenta around $1 \lesssim p \lesssim 2$~fm$^{-1}$, see Figs.~4 and 8 of \ct{Holt:2009ty}. 
With regard to the density-dependent corrections from the 3BF that appear in $NN$ as 
well as in $\Lambda N$ it turns out that they are of comparable order of magnitude. 
For example, 3BF effects driven by two-pion exchange (topology (3)) lead to 
modifications by roughly 40~\% in case of $NN$ and by around 20~\% for $\Lambda N$ at $\rho = \rho_0$
if we take the $S$-wave results at $p = 0$ fm$^{-1}$ as measure. A similar behavior is seen
for the effect of the contact term (topology (6)). Here one has to keep in mind 
that the relevant LEC for $\Lambda N$, $H'$, has only been roughly estimated using scale arguments. 
In particular, as has been demonstrated above, in an EFT that 
includes decuplet baryons as effective degrees of freedom, the 3BF due to a contact 
interaction emerges already at NLO for the $\Lambda NN$ system whereas in case of $NNN$ the 
corresponding 3BF appears only at NNLO. 
In any case, the density-dependent corrections 
in $^1S_0$ and $^3S_1$ due to two-pion exchange and the contact term are of the same sign (repulsive) 
for $NN$ and $\Lambda N$. Those from one-pion exchange are attractive for $NN$ and are likewise
attractive for $\Lambda N$ for the choice of $H'$ being negative. 

%%%%%%%%%%%%%%%%%%%%%%%%%%%%%%

\begin{figure}
\centering
\includegraphics[scale=0.55]{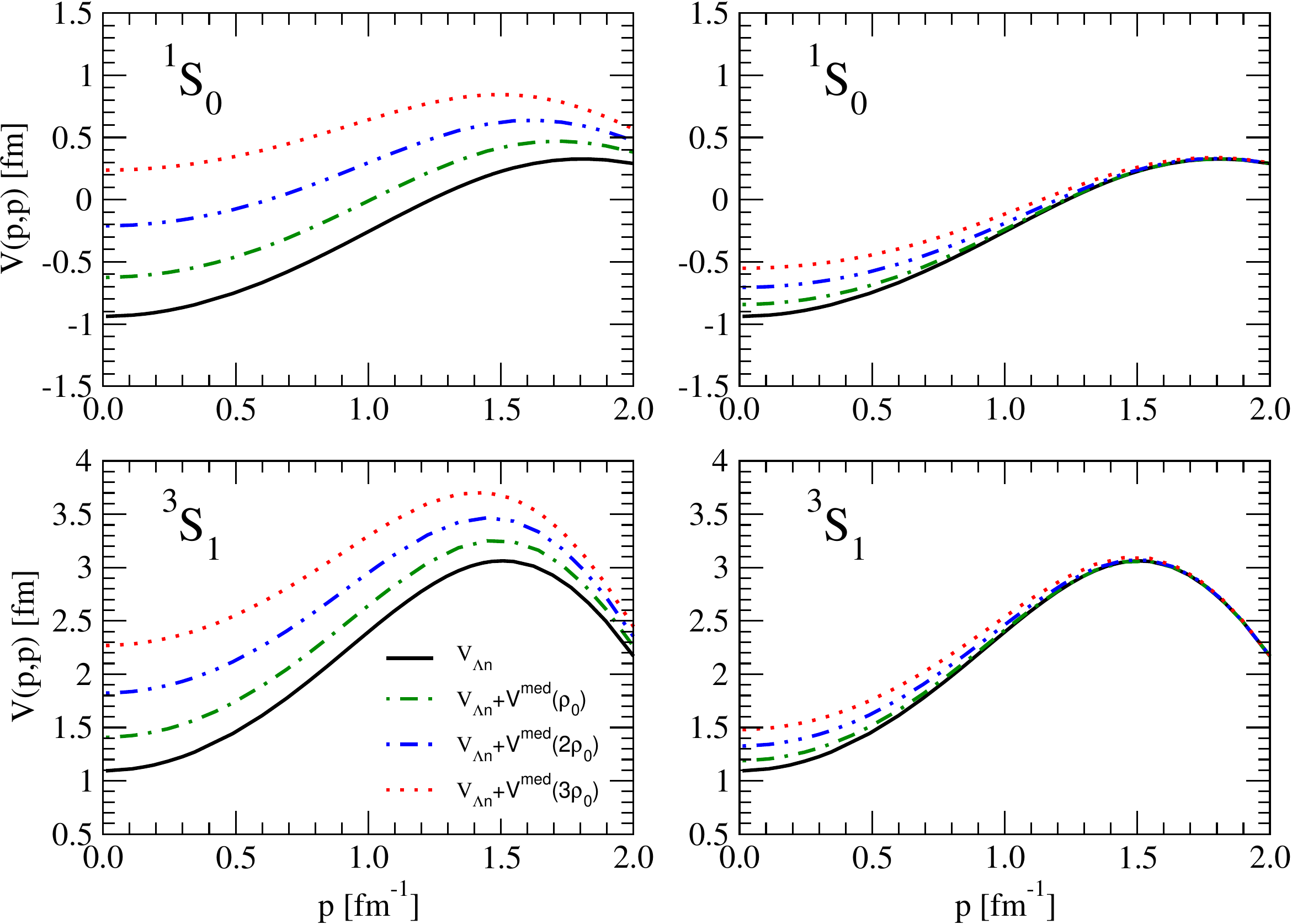}
\vspace{-.5\baselineskip}
\caption{
Modifications of the (on-shell) $\Lambda n$  potential (solid/black line)
due to the combined density-dependent contributions resulting from the 
NLO three-body force. The calculations are for symmetric nuclear matter with 
$\rho = \rho_0$ (dash-dotted/green), $ 2\,\rho_0$ (dash-double dotted/blue), 
and $ 3\,\rho_0$ (dotted/red). 
The $YN$ potential at NLO in chiral EFT with cutoff $\Lambda = 450$ MeV 
\cite{Haidenbauer2013a} is used as basis (black curves).  
The left panel shows results for $H'=+1/f_\pi^2$, the right panel for $H'=-1/f_\pi^2$.
\label{fig:DDs}
}
\vspace{.5\baselineskip}
\includegraphics[scale=0.55]{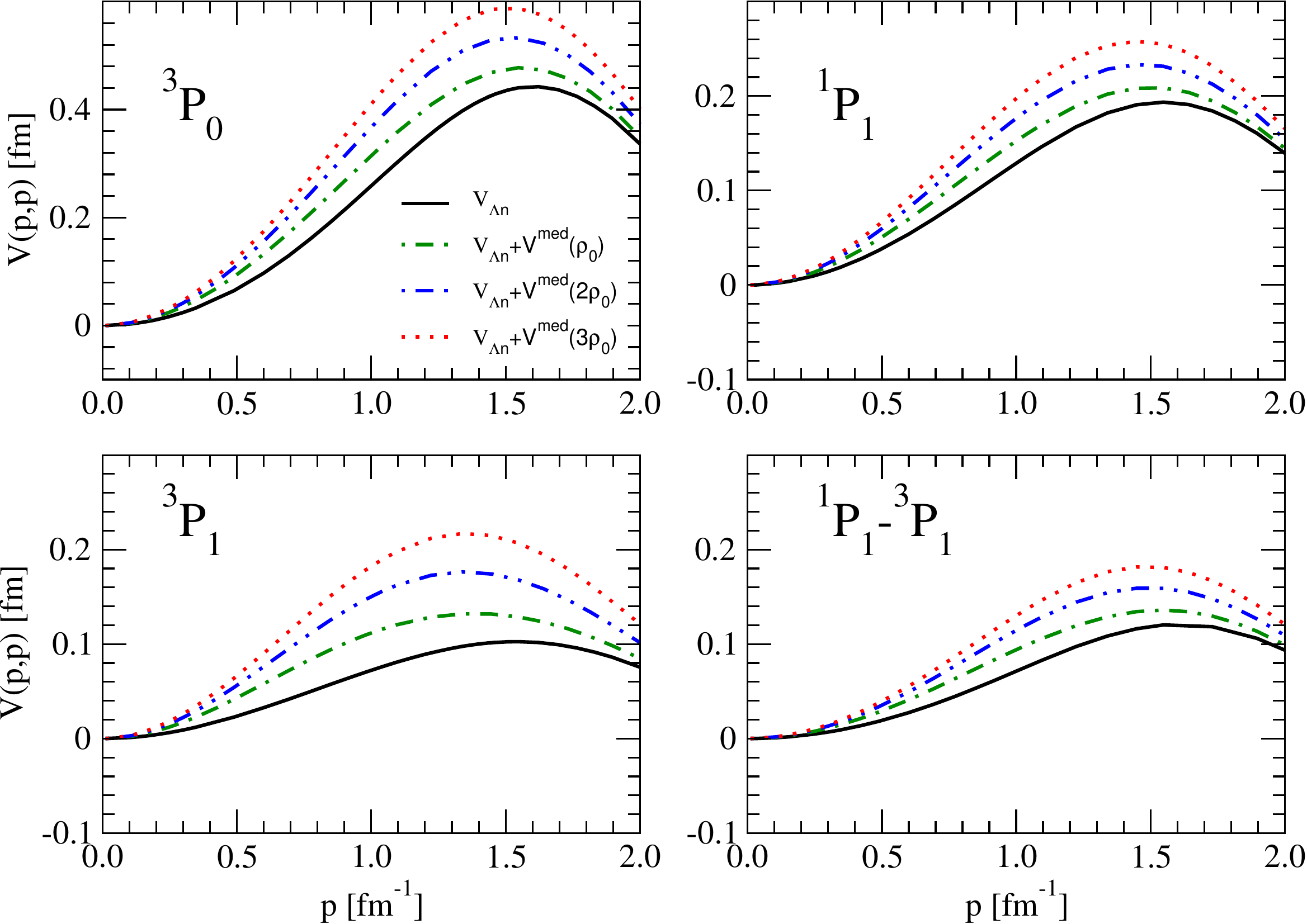}
\vspace{-.5\baselineskip}
\caption{
Modifications of the (on-shell) $\Lambda n$  potential (solid/black line)
due to the combined density-dependent contributions resulting from the 
NLO three-body force for selected \(P\)-waves.
Same description as in Fig.~\ref{fig:DDs}. 
\label{fig:DDp}
}
\end{figure}
Figs.~\ref{fig:DDs} and \ref{fig:DDp} demonstrate how the sum of all terms in the density-dependent
$\Lambda n$ interaction varies with the density in symmetric nuclear matter.
Results for $\rho = \rho_0$, $2\,\rho_0$, and $ 3\,\rho_0$ are displayed. For the choice
$H'=+1/f^2_0$ all three contributions add up and give rise to a sizable density dependence,
see the left-hand parts of Figs.~\ref{fig:DDs} and \ref{fig:DDp}. The density dependence 
is roughly linear in $\rho$ within the considered range. 
For $H'=-1/f^2_0$ there is a destructive interference between the three contributions so
that here the overall density dependence turns out to be more moderate.
%
%%%%%%%%%%%%%%%%%%%%%%%%%%%%%%
%
\begin{figure}
\centering
\includegraphics[scale=0.55]{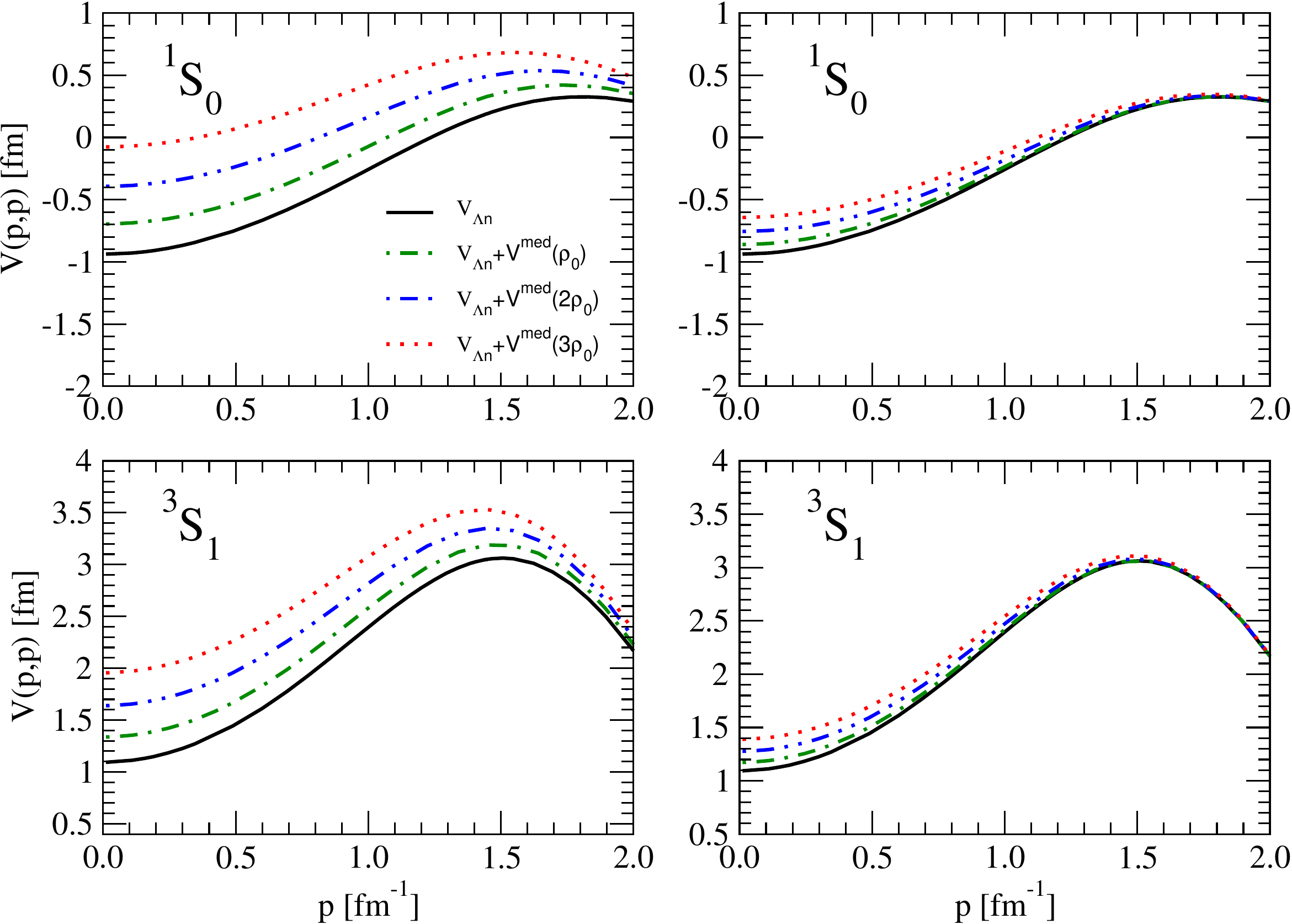}
\vspace{-.5\baselineskip}
\caption{
Modifications of the (on-shell) $\Lambda n$  potential (solid/black line)
due to the combined density-dependent contributions resulting from the 
NLO three-body force.
The calculations are for pure neutron matter with 
$\rho = \rho_0$ (dash-dotted/green), $ 2\,\rho_0$ (dash-double dotted/blue), 
and $ 3\,\rho_0$ (dotted/red). 
The $YN$ potential at NLO in chiral EFT with cutoff $\Lambda = 450$ MeV 
\cite{Haidenbauer2013a} is used as basis (solid curves).  
The left panel shows results for $H'=+1/f_\pi^2$, the right panel for $H'=-1/f_\pi^2$.
\label{fig:DDsN}
}
\vspace{.5\baselineskip}
\includegraphics[scale=0.55]{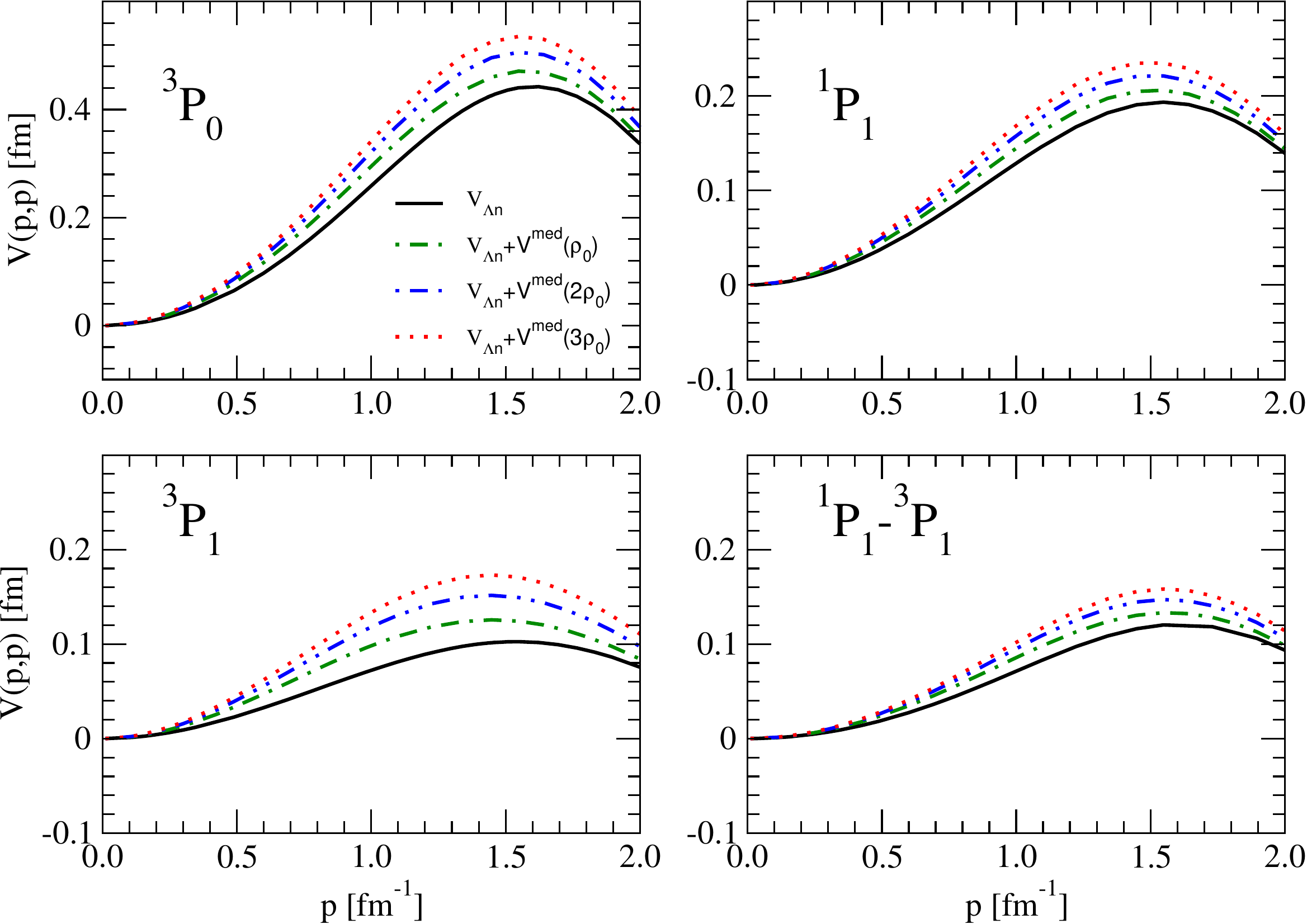}
\vspace{-.5\baselineskip}
\caption{
Modifications of the (on-shell) $\Lambda n$  potential (solid/black line)
due to the combined density-dependent contributions resulting from the 
NLO three-body force for selected \(P\)-waves.
Same description as in Fig.~\ref{fig:DDsN}. 
\label{fig:DDpN}
}
\end{figure}
Finally, in Figs.~\ref{fig:DDsN} and \ref{fig:DDpN} results for pure neutron matter
are presented, again for $\rho = \rho_0$, $2\,\rho_0$, and $ 3\,\rho_0$.
One can see that the resulting density dependence is somewhat smaller than what 
was found for symmetric nuclear matter.
However, one still finds substantial repulsion in neutron matter at moderately high densities.
This result is very encouraging in view of neutron star matter and the hyperon puzzle. In combination with repulsive effects from the momentum-dependent two-body $\Lambda N$ interaction at large Fermi momentum, 
a repulsive hyperon-nucleon-nucleon three-body interaction can potentially play a key role in solving the hyperon puzzle.
It is therefore compelling to employ this effective interaction in neutron star calculations.

\section{Summary and Outlook} \label{sec:sum}

In this work we have presented a systematic construction of density-dependent in-medium two-baryon interactions that follow from the leading chiral three-baryon forces.
These effective potentials should be particularly useful in calculations of many-body systems where an exact treatment of the 
chiral three-baryon forces would otherwise be computationally prohibitive.
Given the systematic derivation of 3BFs within SU(3) chiral effective field theory their application will hopefully shed light on the importance of 3BFs in 
strangeness nuclear physics.
Our derivation is general and applies in particular to the \(S=-1\) sector involving \(\Lambda N\), \(\Sigma N\) coupled channels.
As a concrete example, we have presented explicit expressions for the density-dependent \(\Lambda N\) effective interaction 
which can be implemented in calculations of heavy hypernuclei and (hyper)nuclear matter.
In view of these possible applications we have also supplied the explicit expressions for the in-medium \(\Sigma N\) potentials in isospin-symmetric nuclear matter.

In order to constrain the number of occurring parameters, we have estimated the low-energy constants 
of the leading chiral 3BFs by decuplet saturation. 
The resulting three-body forces, depending only on two free parameters \(\lC_1\) and \(\lC_2\), 
 can be readily employed in investigations of in-medium properties of the
hyperon-nucleon interaction as well as in 
studies of light hypernuclei 
within microscopic approaches like the Faddeev-Yakubovsky formalism \ct*{Nogga2014a} or the no-core shell model \ct*{Wirth2014,Gazda2015}. 

Utilizing these 3BFs we have investigated the medium modification of the \(\Lambda N\) interaction 
induced by chiral \(\Lambda NN\) three-body forces in symmetric nuclear matter and pure neutron matter.
In particular, we have evaluated numerically 
the contributions to the in-medium \(\Lambda n\) potential related to two-pion exchange, 
one-pion exchange and contact terms in the decuplet approximation.
These results indicate a substantial repulsion arising from the \(\Lambda NN\) 3BF at higher densities.
This finding supports scenarios for solving the hyperon puzzle in neutron star matter through strongly
repulsive effects from hyperon-nucleon-nucleon forces.
Another interesting feature is that the medium corrections provide a repulsive contribution to 
the antisymmetric spin-orbit force, as manifested in the \({}^1P_1\)-\({}^3P_1\) transition matrix element.
A sizable antisymmetric spin-orbit force is welcome 
because it can counterbalance the spin-orbit force generated by the basic two-body interaction 
and, thereby, leads to a rather weak $\Lambda$-nucleus spin-orbit potential as 
indicated by hypernuclear spectroscopy.

\begin{acknowledgments}
We thank Andreas~Nogga for useful discussions.
This work is supported in part by DFG and NSFC through funds provided to the Sino-German CRC~110 ``Symmetries and the Emergence of Structure in QCD''.
The work of UGM was also supported by the Chinese Academy of Sciences (CAS) President's International Fellowship Initiative (PIFI) (Grant No. 2015VMA076). 
\end{acknowledgments}

\appendix % \appendix* to avoid letter
% \section{First appendix}

\section{In-medium nucleon-nucleon interaction from chiral three-nucleon forces} \label{app:NN}

Here, we present the results for the effective \(NN\) interaction in isospin-symmetric nuclear matter of density \(\rho = 2k_f^3 / (3\pi^2)\).
The medium corrections from protons and neutrons in the Fermi sea are summed up, and it is advantageous to present the in-medium \(NN\) potential in terms of isospin operators \(\vec\tau_{1,2}\).

The diagram of topology (1) in \fig{fig:med} (Pauli-blocked pion self energy) leads to the following expression:
\begin{equation}
V^\mathrm{med,1,D+E}_{NN} = \frac{g_A^2\rho}{2f_0^4}\bigg[
\vec\tau_1\cdot\vec\tau_2 \frac{\vec\sigma_1\cdot\vec q\ \vec\sigma_2\cdot\vec q}{(\m^2+q^2)^2}(2c_1\m^2+c_3q^2)
-P^{(\sigma)} P^{(\tau)}\vec\tau_1\cdot\vec\tau_2 \frac{\vec\sigma_1\cdot\vec k\ \vec\sigma_2\cdot\vec k}{(\m^2+k^2)^2}(2c_1\m^2+c_3k^2) \bigg]\,,
\end{equation}
where \(P^{(\sigma)} = \frac12(\mathbbm1+\vec\sigma_1\cdot\vec\sigma_2)\) denotes the spin-exchange operator and \(P^{(\tau)} = \frac12(\mathbbm1+\vec\tau_1\cdot\vec\tau_2)\) the isospin-exchange operator.
Note that the second term is the Fierz transform of the first term. This fermion-exchange contribution has not been presented explicitly in \ct{Holt:2009ty}.
The sum of the topologies (2a) and (2b) leads to the result
\begin{align*}
V^\mathrm{med,2a+b,D}_{NN} ={}& \frac{g_A^2}{16\pi^2f_0^4}\vec\tau_1\cdot\vec\tau_2 \frac{\vec\sigma_1\cdot\vec q\ \vec\sigma_2\cdot\vec q}{q^2+\m^2}\Bigg\{
8c_4 \big[ \frac23k_f^3-\m^2\Gamma_0(p,k_f) \big] \\
& -8c_1\m^2 \big[ \Gamma_0(p,k_f)+\Gamma_1(p,k_f) \big]
 -8(c_3+c_4) \Gamma_2(p,k_f) \\
& -4(c_3+c_4)\frac{q^2}2 \big[\Gamma_0(p,k_f)+2\Gamma_1(p,k_f)+\Gamma_3(p,k_f)\big]
\Bigg\} \,, \numberthis
\end{align*}
where the associated exchange part is given by \(V^\mathrm{med,2a+b,E}_{NN}=-P^{(\sigma)} P^{(\tau)} (V^\mathrm{med,2a+b,D}_{NN}\vert_{\vec q\to -\vec k})\).
The diagram of topology~(3) (Pauli-blocked two-pion exchange) leads to the expression
\begin{align*}
V^\mathrm{med,3,D}_{NN} ={}& \frac{g_A^2}{16\pi^2f_0^4}\Bigg\{
- 12 c_1 \m^2 \big[2\Gamma_0(p,k_f)-(q^2+2\m^2)G_0(p,q,k_f)\big] \\ &
- 3 c_3 \big[ \frac83 k_f^3 - 4(q^2+2\m^2)\Gamma_0(p,k_f) - 2q^2 \Gamma_1(p,k_f) + (q^2+2\m^2)^2G_0(p,q,k_f) \big] \\ &
+ 4 c_4 \vec\tau_1\cdot\vec\tau_2\,\big[(G_0(p,q,k_f)+4G_1(p,q,k_f)+4G_3(p,q,k_f)\big] (\vec q\times\vec p\,)\cdot\vec\sigma_1\, (\vec q\times\vec p\,)\cdot\vec\sigma_2 \\ &
+ 4 c_4 \vec\tau_1\cdot\vec\tau_2\, G_2(p,q,k_f) (q^2\vec\sigma_1\cdot\vec\sigma_2-\vec\sigma_1\cdot\vec q\ \vec\sigma_2\cdot\vec q\,) \\ &
+ \frac{\mathrm i}2 (\vec q\times\vec p\,)\cdot(\vec\sigma_1+\vec\sigma_2) \Big[ -24 c_1 \m^2 \Big(G_0(p,q,k_f)+2G_1(p,q,k_f)\Big) \\ &
\qquad -\big( 12 c_3 + 4 c_4 \vec\tau_1\cdot\vec\tau_2\, \big) \frac12 \Big(2\Gamma_0(p,k_f)+2\Gamma_1(p,k_f) \\ &
\qquad -(q^2+2\m^2)\big(G_0(p,q,k_f)+2G_1(p,q,k_f)\big)\Big) \Big]
\Bigg\} \,, \numberthis
\end{align*}
which involves central, spin-spin, tensor, spin-orbit and quadratic spin-orbit components.
The associated exchange part is given by \(V^\mathrm{med,3,E}_{NN}=-P^{(\sigma)} P^{(\tau)} (V^\mathrm{med,3,D}_{NN}\vert_{\vec q\to -\vec k})\).
The in-medium \(NN\) potential due to topology (4) reads:
\begin{equation}
V^\mathrm{med,4,D+E}_{NN} = -\frac{g_A D\rho}{8f_0^2}\bigg[
\vec\tau_1\cdot\vec\tau_2\frac{\vec\sigma_1\cdot\vec q\ \vec\sigma_2\cdot\vec q}{q^2+\m^2}
-P^{(\sigma)} P^{(\tau)} \vec\tau_1\cdot\vec\tau_2 \frac{\vec\sigma_1\cdot\vec k\ \vec\sigma_2\cdot\vec k}{k^2+\m^2}\,
\bigg]\,.
\end{equation}
The sum of the two topologies (5a) and (5b) leads to an (already antisymmetrized) in-medium \(NN\) potential of the form\footnote{
Note that the on-shell relation
\(
\vec\sigma_1\cdot\vec p \ \vec\sigma_2\cdot\vec p + \vec\sigma_1\cdot\vec p^{\,\prime} \ \vec\sigma_2\cdot\vec p^{\,\prime}
= \left( 2p^2-\frac{q^2}2 \right) \vec\sigma_1\cdot\vec\sigma_2
+ \left(1-\frac{2p^2}{q^2}\right) \vec\sigma_1\cdot\vec q\ \vec\sigma_2\cdot\vec q
-\frac2{q^2}\vec\sigma_1\cdot(\vec q\times\vec p\,)\ \vec\sigma_2\cdot(\vec q\times\vec p\,)
\) holds.
}
\begin{align*}
&V^{\mathrm{med},5a+b}_{NN} = \frac{Dg_A}{4\pi^2f_0^2} \bigg\{ 
\Big(\frac34-\frac14 \vec\tau_1\cdot\vec\tau_2 -\frac12 \vec\tau_1\cdot\vec\tau_2\ \vec\sigma_1\cdot\vec\sigma_2\Big)\Big(\frac23k_f^3-\m^2\Gamma_0(p,k_f)\Big) \\ &
\qquad - \frac34(1- \vec\tau_1\cdot\vec\tau_2) \Big[ \frac{\vec\sigma_1\cdot\vec p \ \vec\sigma_2\cdot\vec p + \vec\sigma_1\cdot\vec p^{\,\prime} \ \vec\sigma_2\cdot\vec p^{\,\prime} }2 \big(\Gamma_0(p,k_f)+2\Gamma_1(p,k_f)+\Gamma_3(p,k_f)\big) 
+ \vec\sigma_1\cdot\vec\sigma_2 \Gamma_2(p,k_f) \Big]
\bigg\} \,. \numberthis
\end{align*}
In fact, this result is equal to the antisymmetrized expression of \ct{Holt:2009ty}.
This can be shown by multiplying Eq.~(24) in \ct{Holt:2009ty} with \(1-P^{(\sigma)} P^{(\tau)}\) and by employing the identity:
\(-9\Gamma_2(p,k_f)-3p^2(\Gamma_0(p,k_f)+2\Gamma_1(p,k_f)+\Gamma_3(p,k_f))+2k_f^3-3m^2\Gamma_0(p,k_f)=0\).
Finally, the diagram of topology (6) (contact interaction) leads to the contribution:
\begin{equation}
V^{\mathrm{med},6}_{NN} = -\frac32 E\rho\, (1-P^{(\sigma)} P^{(\tau)})\,.
\end{equation}
This is obviously in agreement with the antisymmetrized expression in Eq.~(25) of \ct{Holt:2009ty}.

In summary all in-medium potentials agree with the antisymmetrized results given in Sec.~III.A. of \ct{Holt:2009ty}.
This consistency serves as a non-trivial check of our calculation, which is based on a different procedure to construct the in-medium potentials.

\section{Construction of the minimal \texorpdfstring{\boldmath $B^*BBB$}{DBBB} Lagrangian} \label{app:DBBB}

In this appendix we present the derivation of the minimal non-relativistic \(B^*BBB\) contact Lagrangian, involving three octet baryons and one decuplet baryon.
An overcomplete set of such contact terms in the non-relativistic limit reads
\begin{equation} \label{eq:LDBBB}
\mathscr{L} = \sum_{\kappa=1}^7 c^\kappa \!\!\! \sum_{\substack{a,b,c,d,\\e,f,g,h,i=1}}^3 \!\!\! \theta^\kappa_{abcdefghi}
\big[
\left(\bar T_{abc}\vec S^\dagger B_{de}\right)\cdot\left(\bar B_{fg}\vec\sigma B_{hi}\right)+
\left(\bar B_{ed}\vec S\, T_{abc}\right)\cdot\left(\bar B_{ih}\vec\sigma B_{gf}\right)
\big] \,,
\end{equation}
with seven different (SU(3) symmetric) flavor structures \(\theta^\kappa\),
\begin{align*}
\theta^1_{abcdefghi} &= \epsilon_{aeg}\delta_{bd}\delta_{ch}\delta_{fi}\,, \\
\theta^2_{abcdefghi} &= \epsilon_{aeg}\delta_{bf}\delta_{ch}\delta_{di}\,, \\
\theta^3_{abcdefghi} &= \epsilon_{aei}\delta_{bd}\delta_{cf}\delta_{hg}\,, \\
\theta^4_{abcdefghi} &= \epsilon_{aei}\delta_{bh}\delta_{cf}\delta_{dg}\,, \\
\theta^5_{abcdefghi} &= \epsilon_{agi}\delta_{bf}\delta_{cd}\delta_{eh}\,, \\
\theta^6_{abcdefghi} &= \epsilon_{agi}\delta_{bh}\delta_{cd}\delta_{ef}\,, \\
\theta^7_{abcdefghi} &= \epsilon_{egi}\delta_{ad}\delta_{bf}\delta_{ch}\,, \numberthis
\end{align*}
and seven associated LECs \(c^\kappa\).
In the particle basis the contact Lagrangian \eq{eq:LDBBB} transforms to
\begin{equation}
\mathscr{L} = \sum_{\kappa=1}^7 c^\kappa \sum_{i,j,k,l} N^\kappa_{B_i^*B_jB_kB_l} 
\big[
\left(\bar B_i^*\vec S^\dagger B_j\right)\cdot\left(\bar B_k\vec\sigma B_l\right)+
\left(\bar B_j\vec S\, B_i^*\right)\cdot\left(\bar B_l\vec\sigma B_k\right)
\big] \,,
\end{equation}
where \(i\) runs now over decuplet baryons, \(j,k,l\) run over octet baryons in the particle basis and the \(N\)'s are certain SU(3) coefficients.

In order to get a minimal set of terms for this contact Lagrangian, we study the processes \(BB\to B^*B\) in more detail.
The corresponding transition matrix elements are derived from the two diagrams
\begin{equation}
V=
\parbox[c][2.3cm][c]{2cm}{\quad
\begin{overpic}[scale=.5]{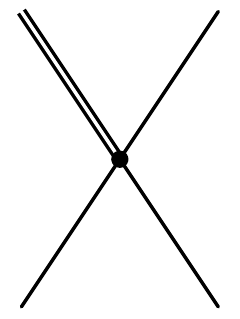}
\put(-10,104){$B^*$}\put(65,104){$B_3$}
\put(-10,-20){$B_1$}\put(65,-20){$B_2$}
\end{overpic}
}
-
\parbox[c][2.3cm][c]{2cm}{\quad
\begin{overpic}[scale=.5]{FvertDBBB}
\put(-10,104){$B^*$}\put(65,104){$B_3$}
\put(-10,-20){$B_2$}\put(65,-20){$B_1$}
\end{overpic}
}
\cdot P^{(\sigma)}
\,,
\end{equation}
where the spin exchange operator \(P^{(\sigma)}=\frac12(\mathbbm{1}+\vec\sigma_1\cdot\vec\sigma_2)\) acts on the initial state.
Making use of the (explicitly verified) identity \(\vec S^\dagger_1\cdot\vec\sigma_2\, P^{(\sigma)} = \vec S^\dagger_1\cdot\vec\sigma_2\), one obtains the following expression for the transition matrix elements:
\begin{equation}
V = -\vec S^\dagger_1\cdot\vec\sigma_2\
\sum_{\kappa=1}^7 c^\kappa \Big(
N^\kappa_{B^*B_{1}B_{3}B_{2}} - N^\kappa_{B^*B_{2}B_{3}B_{1}}
\Big) \,.
\end{equation}
We can obtain the minimal effective Lagrangian by eliminating redundant terms until the rank of the matrix formed by all transitions matches the number of terms in the Lagrangian, as we have done in \ct{Petschauer2016}.
By choosing the two independent flavor structures \(\theta^1\) and \(\theta^2\) one arives at the minimal non-relativistic \(B^*BBB\) Lagrangian written in \eq{eq:LDBBBmin}, with two low-energy constants \(\lC_1\) and \(\lC_2\).
This number of independent constants can be easily understood through group theoretical considerations of the transition \(BB\to B^*B\).
In flavor space the two initial octet baryons form the tensor product \(\mathbf8\otimes\mathbf8\), and in spin space they form the product \(\mathbf{2} \otimes \mathbf{2}\).
These decompose into the irreducible representations as follows:
\begin{equation} \label{eq:BBgroup}
\mathbf8\otimes\mathbf8 = \underbrace{{\mathbf{27}}\oplus{\mathbf{8}_s}\oplus{\mathbf1}}_\text{symmetric}\oplus\underbrace{\mathbf{10}\oplus\mathbf{10^*}\oplus\mathbf{8}_a}_\text{antisymmetric}\,,
\qquad \mathbf{2} \otimes \mathbf{2} = \mathbf{1}_a \oplus \mathbf{3}_s \,.
\end{equation}
Similarly, one finds for the final state with a decuplet and an octet baryon, the following decomposition in flavor and spin space
\begin{equation} \label{eq:DBgroup}
\mathbf{10}\otimes\mathbf8 = \mathbf{35}\oplus\mathbf{27}\oplus\mathbf{10}\oplus\mathbf{8}\,,
\qquad\qquad\qquad\, \mathbf{4} \otimes \mathbf{2} = \mathbf{3} \oplus \mathbf{5} \,.
\end{equation}
At leading order only \(S\)-waves are involved and transitions can only occur between irreducible (flavor and spin) representations of the same type.
This implies, that only transitions in the spin-triplet representation \(\mathbf{3}\) are allowed.
Due to the Pauli principle, the symmetric \(\mathbf{3}\) in spin space must combine with the antisymmetric flavor
representations \(\mathbf{10},\mathbf{10^*},\mathbf{8}_a\) (in the initial state).
Out of these only \(\mathbf{10}\) and \(\mathbf{8}_a\) possess a counterpart in the final state flavor space.
The number of two allowed transitions between irreducible representations corresponds exactly to the number of two LECs in the minimal Lagrangian.
As a consistency check one finds, that the spin-operator \(\vec S_1^{\,\dagger} \cdot \vec\sigma_2\) has a non-vanishing matrix element only for the transition \({}^{3}S_1\rightarrow{}^{3}S_1\).

Another interesting observation can be made from \eqs{eq:BBgroup} and \eq*{eq:DBgroup}.
When restricting to \(NN\) states, only the flavor representations \(\mathbf{27}\) and \(\mathbf{10^*}\) are involved (cf.\ for example \ct{Polinder2006}).
But these representations combine either with the wrong total spin, or have no counterpart in the final state.
Hence, \(NN\to \Delta N\) transitions in \(S\)-waves are forbidden due to the Pauli principle.

% \flushleft
 %\bibliography{../../Mendeley/BibTeX/zPaperBib-20163BFMatter}
 %\bibliography{../../../../../Bücher/Mendeley/BibTeX/zPaperBib-20163BFMatter}
 \input{bib.tex}

%\begin{thebibliography}{999}
%...
%\end{thebibliography}

\end{document}

%% file: files/BBMBBDecconstants.tex
\ld_1  &= -\frac{7 C  (\lC_1 +\lC_2 )}{18 \Delta } \,, \\ 
\ld_2  &= -\frac{C  (\lC_1 -7 \lC_2 )}{18 \Delta } \,, \\ 
\ld_3  &= \frac{C  (3 \lC_1 +11 \lC_2 )}{18 \Delta } \,, \\ 
\ld_4  &= -\frac{C  (9 \lC_1 +13 \lC_2 )}{18 \Delta } \,, \\ 
\ld_5  &= -\frac{C  (\lC_1 -3 \lC_2 )}{18 \Delta } \,, \\ 
\ld_6  &= -\frac{C  (5 \lC_1 -3 \lC_2 )}{18 \Delta } \,, \\ 
\ld_7  &= \frac{2 \lC_2  C }{9 \Delta } \,, \\
\ld_8  &= -\frac{C  (5 \lC_1 -3 \lC_2 )}{18 \Delta } \,, \\
\ld_9  &= -\frac{C  (5 \lC_1 +9 \lC_2 )}{9 \Delta } \,, \\ 
\ld_{10}  &= -\frac{5 C  (\lC_1 +\lC_2 )}{6 \Delta } \,, \\ 
\ld_{11}  &= \frac{C  (\lC_1 +9 \lC_2 )}{18 \Delta } \,, \\ 
\ld_{12}  &= \frac{C  (2 \lC_1 +5 \lC_2 )}{9 \Delta } \,, \\ 
\ld_{13}  &= \frac{C  (\lC_1 +5 \lC_2 )}{18 \Delta } \,, \\
\ld_{14}  &= -\frac{C  (3 \lC_1 +7 \lC_2 )}{18 \Delta } \,. \numberthis

%% file: files/BBBDecconstants.tex
\lc_1  &= -\frac{7 (\lC_1 +\lC_2 )^2}{24 \Delta } \,, \\ 
\lc_2  &= -\frac{\lC_1 ^2+18 \lC_1  \lC_2 +9 \lC_2 ^2}{36 \Delta } \,, \\ 
\lc_3  &= -\frac{19 \lC_1 ^2+30 \lC_1  \lC_2 +15 \lC_2 ^2}{36 \Delta } \,, \\ 
\lc_4  &= \frac{\lC_1 ^2+18 \lC_1  \lC_2 +9 \lC_2 ^2}{72 \Delta } \,, \\ 
\lc_5  &= \frac{5 (\lC_1 +\lC_2 )^2}{8 \Delta } \,, \\ 
\lc_6  &= \frac{17 \lC_1 ^2+18 \lC_1  \lC_2 -15 \lC_2 ^2}{72 \Delta } \,, \\ 
\lc_7  &= \frac{7 \lC_1 ^2+6 \lC_1  \lC_2 -9 \lC_2 ^2}{108 \Delta } \,, \\ 
\lc_8  &= \frac{25 \lC_1 ^2+42 \lC_1  \lC_2 -3 \lC_2 ^2}{108 \Delta } \,,\rule{0pt}{20pt} \\  %!!!!
\lc_9  &= \frac{\lC_1 ^2+18 \lC_1  \lC_2 +9 \lC_2 ^2}{72 \Delta } \,, \\ 
\lc_{10}  &= -\frac{25 \lC_1 ^2+50 \lC_1  \lC_2 +9 \lC_2 ^2}{72 \Delta } \,, \\ 
\lc_{11}  &= -\frac{23 (\lC_1 +\lC_2 )^2}{72 \Delta } \,, \\ 
\lc_{12}  &= -\frac{13 \lC_1 ^2+42 \lC_1  \lC_2 +21 \lC_2 ^2}{108 \Delta } \,, \\
\lc_{13}  &= -\frac{\lC_1 ^2+10 \lC_1  \lC_2 +5 \lC_2 ^2}{36 \Delta } \,,\rule{0pt}{23pt} \\  %!!!!
\lc_{14}  &= \frac{5 (\lC_1 +\lC_2 )^2}{24 \Delta } \,, \\ 
\lc_{15}  &= -\frac{\lC_1 ^2-9 \lC_2 ^2}{27 \Delta } \,, \\ 
\lc_{16}  &= -\frac{11 \lC_1 ^2+18 \lC_1  \lC_2 +3 \lC_2 ^2}{54 \Delta } \,, \\ 
\lc_{17}  &= -\frac{2 \lC_1  (\lC_1 +2 \lC_2 )}{9 \Delta } \,, \\ 
\lc_{18}  &= \frac{2 \lC_1 ^2}{27 \Delta } \,. \numberthis